\documentclass[11pt]{article}
\pdfoutput=1
\usepackage{jcappub}

\usepackage{color}
\input{colordvi.tex}

\usepackage{graphicx}
\usepackage{float}
\usepackage{wrapfig}
\usepackage{bm}
\usepackage{amsmath,amssymb}
\usepackage{subfigure}
\usepackage{verbatim}
\usepackage{makeidx}

\allowdisplaybreaks[4]

\title{CMB spectral distortions as solutions to the Boltzmann equations}
\author{Atsuhisa Ota}
\affiliation{Department of Physics, Tokyo Institute of Technology,\\
Tokyo 152-8551, Japan}
\emailAdd{a.ota@th.phys.titech.ac.jp}

\abstract{
Distortions to the cosmic microwave background~(CMB) blackbody spectrum are calculated in the framework of cosmological perturbation theory.
The second order Boltzmann equation is explicitly solved, with the spectral $y$ distortion and the frequency independent second order temperature perturbation.
We also solve higher order Boltzmann equations systematically and find new type spectral distortions in the low energy limit of electrons.
As an example, we concretely construct a solution to the cubic order Boltzmann equation and show that it can be characterized by three parameters: a cubic order temperature perturbation and two different types of cubic order spectral distortions.
A new linear Sunyaev-Zel'dovich effect whose frequency dependence is different from the usual $y$ distortion is also discussed in the presence of the next-to-leading order Kompaneets terms, and we show that higher order spectral distortions are also generated as a result of the diffusion process in a framework of the higher order cosmological perturbation theory.
We also comment that generation of the spectral $\mu$ distortion cannot be explained in this framework even at second order.
}


\begin{document}
\maketitle

\section{Introduction}

The temperature anisotropy analysis of the cosmic microwave background~(CMB) has been successful in the past few decades.
The standard $\Lambda$CDM cosmology explains the current linear fluctuations very well and shows us the efficiency of inflationary paradigm.
One of the remaining issues of modern cosmology is to identify the concrete model of the early universe.
The CMB temperature bispectrum has been intensely investigated in this context, and the Planck satellite provides us tight constraints on the primordial non-Gaussianity~\cite{Ade:2015ava}.
Towards future observations, the second order Boltzmann theory has been developed~\cite{Bartolo:2006cu,Pitrou:2007jy,Pitrou:2008ut,Beneke:2010eg,Naruko:2013aaa} because secondary effects may mimic the primordial non-Gaussianity~\cite{Bartolo:2003bz,Bartolo:2004ty,Bartolo:2011wb,Senatore:2008wk,Khatri:2008kb,Nitta:2009jp,Creminelli:2004pv,Boubekeur:2009uk,Lewis:2012tc,Creminelli:2011sq,Fidler:2014zwa}.
The smallness of the principle anisotropies makes it crucial to estimate the secondary anisotropies, and nowadays several numerical codes have been developed to get rid of such contaminations~\cite{Pitrou:2010sn,Su:2012gt,Huang:2012ub,Pettinari:2013he,Huang:2013qua,Saito:2014bxa}.
The anisotropies in the $y$ distortions are also discussed in the same context~\cite{Pitrou:2009bc,Renaux-Petel:2013zwa}.

On the other hand, distortions to the CMB blackbody spectrum have also been discussed to investigate the thermal history of the early universe.
One example is the observed thermal Sunyaev-Zel'dovich (tSZ) effect due to hot gas in the intracluster medium of galaxy groups~\cite{Zeldovich:1969ff,Sunyaev:1970er,Aghanim:2015eva}.
More generally, the spectral distortions are powerful tools to investigate energy injections from several non-trivial processes such as dark matter pair annihilation~\cite{Hu:1993gc,McDonald:2000bk,Chluba:2009uv}, evaporation of the primordial blackholes~\cite{Carr:2009jm} and dissipations of the primordial fluctuations~\cite{Sunyaev:1970eu,Hu:1994bz,Chluba:2011hw,1991MNRAS.248...52B,Chluba:2012gq,Chluba:2012we,Khatri:2012rt,Khatri:2012tw,Khatri:2013dha,Clesse:2014pna,Ota:2014hha,Chluba:2014qia}.
Recently, anisotropies of the distortions from Silk damping are also discussed for a primordial non-Gaussianity observation~\cite{Pajer:2012vz,Pajer:2013oca,Ganc:2012ae,Ganc:2014wia,Ota:2014iva,Naruko:2015pva,Ota:2016mqd,Emami:2015xqa,Chluba:2016aln}.
The previous analysis is based on an assumption of local thermodynamics, which is not the case for the late universe before recombination. 
These anisotropies are at second order in the primordial curvature perturbations and can be understood as mode coupling effects in a framework of the second order Boltzmann theory.

In this paper, we give a unified view of the above two issues by explicitly solving the Boltzmann equations with momentum~(frequency) dependent temperature perturbations.
We discuss frequency dependence of the temperature perturbations which arises at nonlinear order.
There is an infinite number of evolution equations corresponding to each frequency.
As pointed out, for example, in~\cite{Dodelson:1993xz}, we usually integrate the photon momentum and define the brightness perturbations at second order.  
This simplification has been widespread since the anisotropy experiments do not focus on spectroscopy; however, we should keep in mind that much information is hidden in the nontrivial momentum dependence.
We handle such an infinite number of d.o.f. coming from the continuous momentum by replacing it with an infinite number of the parameters describing spectral distortions~\footnote{A similar manipulation called \textit{moment expansion} was initially proposed in~\cite{Stebbins:2007ve,Pitrou:2014ota}.
The authors also tried to classify the spectral distortions systematically.
In their literature, detailed analyses were not discussed in the presence of second or higher order collision terms, which are crucial for spectral distortions and central topics of this paper.}.
Then, we find that the second order photon Boltzmann equation falls into two frequency independent equations.
Besides, we point out a possibility to solve the higher order Boltzmann equations systematically by introducing higher order spectral distortions.
As an example, we construct the cubic order spectral distortions in our method.
Observing such a little higher order spectral distortion can be the future works in the next few decades.
On the other hand, the method may be useful as a prescription to non-equilibrium physics or non-linear evolution of the large-scale structure~(LSS) as fluid dynamics of matters with gravitational interactions.

We organize this paper as follows.
In section \ref{sec:2}, we summarize second and third order Boltzmann collision terms for the Compton scattering~\footnote{Section~\ref{sec:2} is considerably revised.}.
The ansatz and the coefficient equations are discussed in section \ref{sec:3}.
We solve the equation for the $y$ distortion in section \ref{sec:4} and confirm availability of the previous phenomenological estimations.
In section \ref{sec:5}, we comment on another spectral distortion called $\mu$ distortion.
Section \ref{sec:6} is devoted to the extension of the method to higher orders cosmological perturbations.
The appendices provide several definitions and comparisons our results with the previous works.
We then conclude in the final section.

\section{Boltzmann equation}\label{sec:2}

We begin with deriving the higher order Boltzmann collision terms to investigate the spectral distortions and higher order temperature anisotropies.

\subsection{Set up}

Let $f$ and $g$ be the distribution functions for photons and electrons, respectively.
Ignoring the Pauli blocking factors, the Boltzmann collision term is given as

\begin{align}
\mathcal C[f]=&\frac{1}{16\pi }\int
\frac{d\mathbf n'}{4\pi}d{\tilde p}'\int \frac{d^3\tilde q}{(2\pi)^3}\frac{1}{4}\sum_{\rm spins}|\mathcal M|^2\notag \\
&\times \frac{{\tilde p}'}{{\tilde p}}\frac  1{E(\mathbf {\tilde q})E(\mathbf {\tilde p}+\mathbf {\tilde q}-\mathbf {\tilde p}')}\delta[{\tilde p}+E(\mathbf {\tilde q})-{\tilde p}'-E(\mathbf {\tilde p}+\mathbf {\tilde q}-\mathbf {\tilde p}')]\notag \\
&\times \left\{g(\mathbf {\tilde q}')f(\mathbf {\tilde p}')[1+f(\mathbf {\tilde p})]-g(\mathbf {\tilde q})f(\mathbf {\tilde p})[1+f(\mathbf
{\tilde p}')]\right\},\label{Col:expand}
\end{align}
where tildes imply that they are physical momenta, which are different from the comoving momenta without tildes~(e.g. $p(1+z)=\tilde p$ with the redshift $z$).
$\mathcal M$ is the invariant scattering amplitude and $\mathbf n'\equiv \mathbf p'/|\mathbf p'|$.
Let us expand each part of (\ref{Col:expand}) by introducing two parameters: $\epsilon\sim \mathcal O(q/m_{\rm e})$ and $\eta=\mathcal O(|\mathbf {\tilde p}-\mathbf {\tilde p}'|/m_{\rm e})$~\footnote{We should note that there are two types hierarchies: the cosmological perturbations and the electron energy transfer.
The former is directly related to the primordial quantum fluctuations, and we usually consider that the magnitude is the order of $10^{-5}$ at first order if we assume their scale invariance.
In this paper, we use the terminologies ``first order'' and ``second order'' regarding the cosmological perturbations. 
}.
We consider that $\epsilon$ is the order of the electron bulk velocity $|\mathbf v|=\mathcal O(10^{-5})$ or thermal motion $\sqrt{T_{\rm e}/m_{\rm e}}$. For the latter, the relation between the two parameters are given as $\epsilon^2 \sim \eta\ll 1$ since the photon number has a peak around ${\tilde p}\sim T_\gamma$, and $T_\gamma\sim T_{\rm e}$ is manifest if we assume that the system is in kinetic equilibrium~\footnote{In the hot electron gas in the late universe, $\epsilon^2\gg \eta$ is possible.}
~\footnote{We would like to thank Taku Haga and Keisuke Inomata for pointing out misprints in the previous manuscripts.}.\\

\quad

\textit{The Scattering amplitude---.}
Let us denote the initial state 4-momenta with primes.
The scattering amplitude is then written as~\cite{Peskin:1995ev}
\begin{align}
\frac{1}{4}\sum_{\rm spins}|\mathcal M|^2=2{\rm e}^4\left[\frac{{\tilde q}\cdot {\tilde p}'}{{\tilde q}\cdot {\tilde p}}+\frac{{\tilde q}\cdot {\tilde p}}{{\tilde q}\cdot {\tilde p}'}-2m_{\rm e}^2\left(\frac{1}{{\tilde q}\cdot {\tilde p}}-\frac{1}{{\tilde q}\cdot {\tilde p}'}\right)+m_{\rm e}^4\left(\frac{1}{{\tilde q}\cdot {\tilde p}}-\frac{1}{{\tilde q}\cdot {\tilde p}'}\right)^2\right],\label{KN:first}
\end{align}
where we have used
\begin{align}
{\tilde p}\cdot {\tilde q}&={\tilde p}'\cdot {\tilde q}',\\
{\tilde p}\cdot {\tilde q}'&={\tilde p}'\cdot {\tilde q},
\end{align}
which are obtained with 4-momentum conservation ${\tilde p}+{\tilde q}={\tilde p}'+{\tilde q}'$.
Let us chose a frame satisfying ${\tilde q}_\mu=(-\sqrt{m^2_{\rm e}+\mathbf {\tilde q}^2},\mathbf {\tilde q})$.
Using the above, we can expand (\ref{KN:first}) as follows: 
\begin{align}
\frac{1}{4}\sum_{\rm spins}|\mathcal M|^2=|\mathcal M_0|^2+|\mathcal M_\epsilon |^2+|\mathcal M_{\eta^2}|^2 + |\mathcal M_{\epsilon^2}|^2+ |\mathcal M_{\epsilon^3}|^2,\label{KN:expand}
\end{align}
where we have defined
\begin{align}
\frac{|\mathcal M_0|^2}{2{\rm e}^4} &=1+\lambda^2, \\
\frac{|\mathcal M_\epsilon |^2}{2{\rm e}^4} &=2\lambda (\lambda-1)\left[\frac{\mathbf n\cdot \mathbf {\tilde q}}{m_{\rm e}}+\frac{\mathbf n'\cdot \mathbf {\tilde q}}{m_{\rm e}}\right], \\
\frac{|\mathcal M_{\eta^2}|^2}{2{\rm e}^4}  &=(\lambda-1)^2\frac{{\tilde p}^2}{{m^2_{\rm e}}},\\
\frac{|\mathcal M_{\epsilon^2}|^2}{2{\rm e}^4} &=-2\lambda(\lambda-1)\frac{{\tilde q}^2}{{m^2_{\rm e}}}+
(\lambda -1) (3 \lambda -1)
\left[\frac{(\mathbf n'\cdot \mathbf  {\tilde q})^2}{{m^2_{\rm e}}}+  \frac{(\mathbf n\cdot \mathbf  {\tilde q})^2}{{m^2_{\rm e}}}\right] \notag \\
&+2 (\lambda -1) (2 \lambda -1) \frac{(\mathbf n\cdot \mathbf {\tilde q})(\mathbf n'\cdot \mathbf {\tilde q})}{{m^2_{\rm e}}}.
\end{align}

\textit{The energy product---.}
On the other hand, the product of energy can be reduced to as follows:

\begin{align}
\frac{{\tilde p}'}{{\tilde p}}\frac{1}{E(\mathbf {\tilde q})E(\mathbf {\tilde q}')}
= \frac{{\tilde p}'}{
{\tilde p}m_{\rm e}^2}\left(1-\mathcal E_{\epsilon^2}-\mathcal E_{\epsilon \eta}-\mathcal E_{\eta^2} \right) \label{En:expand},
\end{align}
where each part is written as
\begin{align}
\mathcal E_{\epsilon^2}&=\frac{{\tilde q}^2}{{m^2_{\rm e}}},\\
\mathcal E_{\epsilon \eta }&=\frac{(\mathbf {\tilde p}-\mathbf {\tilde p}')\cdot \mathbf {\tilde q}}{{m^2_{\rm e}}},\\
\mathcal E_{\eta^2}&=\frac{(\mathbf {\tilde p}-\mathbf {\tilde p}')^2}{2{m^2_{\rm e}}}.
\end{align}

\textit{The delta function---.}
The delta functions are Taylor expanded around the photon energy difference; that is, we have
\begin{align}
&\delta({\tilde p}-{\tilde p}'+E(\mathbf {\tilde q})-E(\mathbf {\tilde q}'))\notag \\
&=\delta({\tilde p}-{\tilde p}')+\left(\mathcal D_{\epsilon}+\mathcal D_{\epsilon^3}+\mathcal D_{\eta}\right)\frac{\partial \delta({\tilde p}-{\tilde p}')}{\partial {\tilde p}'}\notag \\
&+\frac12\left(\mathcal D_{\epsilon}+\mathcal D_{\eta}\right)^2\frac{\partial^2 \delta({\tilde p}-{\tilde p}')}{\partial {\tilde p}'^2}+\frac{1}{3!}\mathcal D^3_{\epsilon}\frac{\partial^{3} \delta({\tilde p}-{\tilde p}')}{\partial {\tilde p}'^{3}}\label{delta:expand},
\end{align}
where we have introduced
\begin{align}
\mathcal D_\epsilon &=\frac{(\mathbf {\tilde p}-\mathbf {\tilde p}')\cdot \mathbf {\tilde q}}{m_{\rm e}},\\
\mathcal D_{\epsilon^3} &=-\frac{{\tilde q}^2(\mathbf {\tilde p}-\mathbf {\tilde p}')\cdot \mathbf {\tilde q}}{2m_{\rm e}^3},\\
\mathcal D_\eta &=\frac{(\mathbf {\tilde p}-\mathbf {\tilde p}')^2}{2m_{\rm e}}.
\end{align}

\textit{Product of distribution functions---.}
Expanding $g(\mathbf {\tilde q}')$ around $\mathbf {\tilde q}$, the product of the distribution functions can be written as 

\begin{align}
\mathcal F(\mathbf {\tilde p}',\mathbf {\tilde p})&=\frac{g(\mathbf {\tilde q'})}{g(\mathbf {\tilde q})} f(\mathbf {\tilde p}')[1+f(\mathbf {\tilde p})]-f(\mathbf {\tilde p})[1+f(\mathbf
{\tilde p}')]\notag \\
&=\mathcal F_1(\mathbf {\tilde p}',\mathbf {\tilde p})+\left(
\alpha_{\eta/\epsilon}+\alpha_{\eta^2/\epsilon^2}
\right)\mathcal F_3(\mathbf {\tilde p}',\mathbf {\tilde p})\label{dist:expand},
\end{align}
where $\mathcal F_1$, $\mathcal F_3$ and $\alpha$'s are defined as
\begin{align}
\alpha_{\eta/\epsilon}&=-\frac{(\mathbf {\tilde q}-m_{\rm e} \mathbf v)\cdot (\mathbf {\tilde p}-\mathbf {\tilde p}')}{m_{\rm e}T_{\rm e}},\\
\alpha_{\eta^2/\epsilon^2}&=-\frac{(\mathbf {\tilde p}-\mathbf {\tilde p}')^2}{2m_{\rm e}T_{\rm e}}+\frac12\alpha^{2}_{\eta/\epsilon},\\
\mathcal F_1(\mathbf {\tilde p}',\mathbf {\tilde p})&=f(\mathbf {\tilde p}')-f(\mathbf {\tilde p}),\\
\mathcal F_3(\mathbf {\tilde p}',\mathbf {\tilde p})&=f(\mathbf {\tilde p}')\left[1+f(\mathbf {\tilde p})\right].
\end{align}

\textit{Electron momentum integrals---.}
Let us write the $\mathbf {\tilde q}$ integral as follows:
\begin{align}
\langle    \cdots \rangle&\equiv \int\frac{d^3 {\tilde q}}{(2\pi)^3}\cdots g(\mathbf {\tilde q}).
\end{align}

Then integrals with the momentum are recast into

\begin{align}
\begin{split}
\langle    {\tilde q}_i\rangle&=n_{\rm e}m_{\rm e}v_i,\\
\langle    {\tilde q}_i{\tilde q}_j\rangle&=n_{\rm e}m_{\rm e} T_{\rm e} \delta_{ij}+n_{\rm e}{m^2_{\rm e}}_{\rm e}v_iv_j,\\
\langle {\tilde q}_i{\tilde q}_j{\tilde q}_k\rangle&=n_{\rm e}m_{\rm e}\left(v_i\langle ({\tilde q}_j-m_{\rm e} v_j)({\tilde q}_k-m_{\rm e}v_k)\rangle+2{\rm perms.}\right)+n_{\rm e}m^3_{\rm e}v_iv_jv_k \\
&=n_{\rm e}{m^2_{\rm e}}T_{\rm e}(v_i  \delta_{jk}+2{\rm perms.})+n_{\rm e}m^3_{\rm e}v_iv_jv_k.\label{electron:int}
\end{split}
\end{align}\\

\subsection{Expansions in terms of $\epsilon$ and $\eta$}
We are ready to expand the collision terms with respect to $\epsilon$ and $\eta$:
\begin{align}
\mathcal C[f]=\mathcal C_{0,0}[f]+\mathcal C_{\epsilon,0}[f]+\mathcal C_{\epsilon^2,0}[f]+\mathcal C_{\epsilon^3,0}[f]+\mathcal C_{0,\eta}[f]+\mathcal C_{\epsilon,\eta}[f].
\end{align}
We substitute (\ref{KN:expand}), (\ref{En:expand}), (\ref{delta:expand}) and (\ref{dist:expand}) into (\ref{Col:expand}).
Then, we integrate each momentum by using (\ref{electron:int}) and substitute ${\tilde p}=p(1+z)$.
To simplify the notation, we introduce the following three parameters: $\lambda = \mathbf n\cdot \mathbf n'$, $V = \mathbf {v}\cdot \mathbf n$ and $V' = \mathbf { v}\cdot \mathbf n'$.
After tremendous but straight forward calculations, each term is obtained as follows:

\begin{align}
(n_{\rm e}\sigma_{\rm T}a)^{-1}\mathcal C_{0,0}[f]&=\int \frac{d\mathbf n'}{4\pi}\frac{3}{4} \left(\lambda ^2+1\right) \mathcal{F}_1(p\mathbf n',p\mathbf n),\\
(n_{\rm e}\sigma_{\rm T}a)^{-1}\mathcal C_{\epsilon,0}[f]&
=\int \frac{d\mathbf n'}{4\pi} \frac{3}{4}  \Bigg[\left\{\left(4 \lambda ^2-2 \lambda +2\right) V '+\left(\lambda ^2-2 \lambda -1\right) V \right\} \mathcal{F}_1(p\mathbf n',p\mathbf n)\notag \\
&\left.
-\left(\lambda ^2+1\right) \left(V -V '\right) p\frac{\partial \mathcal{F}_1(p'\mathbf n',p\mathbf n)}{\partial p'}\bigg|_{p'=p}\right],\\
(n_{\rm e}\sigma_{\rm T}a)^{-1}
\mathcal C_{0,\eta}[f] 
&=\int \frac{d\mathbf n'}{4\pi}\frac{3p(1+z)}{4m_{\rm e}} \left(\lambda ^3-\lambda ^2+\lambda -1\right) \notag\\
&\times \left[p \frac{\partial \mathcal{F}_1(p'\mathbf n',p\mathbf n)}{\partial p'}\bigg|_{p'=p}-2p \frac{\partial \mathcal{F}_3(p'\mathbf n',p\mathbf n)}{\partial p'{}}\bigg|_{p'=p}\right]\notag \\
&+\int \frac{d\mathbf n'}{4\pi}\frac{3p(1+z)}{4 m_{\rm e}} \left(\lambda ^3-\lambda ^2+\lambda -1\right)  \left[2 \mathcal{F}_1(p\mathbf n',p\mathbf n)-4 \mathcal{F}_3(p\mathbf n',p\mathbf n)\right],\\
(n_{\rm e}\sigma_{\rm T}a)^{-1}\mathcal C_{\epsilon,\eta}[f] 
&=\int \frac{d\mathbf n'}{4\pi}\frac{p(1+z)}{m_{\rm e}} \Bigg[\frac{3}{4} (\lambda -1) \left(-\lambda ^2-1\right)  \left(V -V '\right) \notag\\
&\times \left[p^2\frac{\partial^2 \mathcal{F}_1(p'\mathbf n',p\mathbf n)}{\partial p'{}^2}\bigg|_{p'=p}-2 p^2\frac{\partial^2 \mathcal{F}_3(p'\mathbf n',p\mathbf n)}{\partial p'{}^2}\bigg|_{p'=p}\right]\notag \\
&+\frac{3}{4} (\lambda -1) \left[4 \left\{\left(-4 \lambda ^2+\lambda -3\right) V '+\left(\lambda ^2+\lambda +2\right) V \right\}p \frac{\partial \mathcal{F}_3(p'\mathbf n',p\mathbf n)}{\partial p'{}}\bigg|_{p'=p}
\right.\notag \\
&\left.
-2 \left\{\left(-4 \lambda ^2+\lambda -3\right) V '+\left(\lambda ^2+\lambda +2\right) V \right\} p\frac{\partial \mathcal{F}_1(p'\mathbf n',p\mathbf n)}{\partial p'}\bigg|_{p'=p}\right]\notag \\
&+\frac{3}{4} (\lambda -1)  \left[2 \left\{\left(5 \lambda ^2-2 \lambda +3\right) V '+\left(\lambda ^2-2 \lambda -1\right) V \right\} \mathcal{F}_1(p\mathbf n',p\mathbf n)\right.\notag \\
&\left.-4 \left\{\left(5 \lambda ^2-2 \lambda +3\right) V '+\left(\lambda ^2-2 \lambda -1\right) V \right\} \mathcal{F}_3(p\mathbf n',p\mathbf n)\right]\Bigg],
\end{align}

\begin{align}
(n_{\rm e}\sigma_{\rm T}a)^{-1}\mathcal C_{\epsilon^2,0}[f]=&\int \frac{d\mathbf n'}{4\pi} \frac{T_{\rm e}}{m_{\rm e}} \left[-\frac{3}{4}\left(\lambda ^3-\lambda ^2+\lambda -1\right) p^2 \frac{\partial^2 \mathcal{F}_1(p'\mathbf n',p\mathbf n)}{\partial p'^2}\bigg|_{p'=p}
\right.\notag \\
&\left.-3 \left(\lambda ^3-\lambda ^2+\lambda -1\right) p \frac{\partial \mathcal{F}_1(p'\mathbf n',p\mathbf n)}{\partial p'}\bigg|_{p'=p}
\right.\notag \\
&\left.+\frac{3}{2} \left(2 \lambda ^3-3 \lambda ^2-2 \lambda +1\right) \mathcal{F}_1(p\mathbf n',p\mathbf n)\right]\notag 
\\
&+
\int \frac{d\mathbf n'}{4\pi} \bigg[\frac{3}{8} \left(\lambda ^2+1\right) V^2 p^2 \frac{\partial^2 \mathcal{F}_1(p'\mathbf n',p\mathbf n)}{\partial p'^2}\bigg|_{p'=p}
\notag\\
&
+\frac{3}{8} \left(\lambda ^2+1\right)  V '{}^2 p^2 \frac{\partial^2 \mathcal{F}_1(p'\mathbf n',p\mathbf n)}{\partial p'^2}\bigg|_{p'=p}\notag 
\\
&
-\frac{3}{4} \left(\lambda ^2+1\right) V  V ' p^2\frac{\partial^2 \mathcal{F}_1(p'\mathbf n',p\mathbf n)}{\partial p'^2}\bigg|_{p'=p}\notag \\
&-\frac{3}{4} \left(\lambda ^2-2 \lambda -1\right) V ^2 p \frac{\partial \mathcal{F}_1(p'\mathbf n',p\mathbf n)}{\partial p'}\bigg|_{p'=p}\notag 
\\
&
-\frac{3}{4} \left(-5 \lambda ^2+2 \lambda -3\right) V '{}^2 p\frac{\partial \mathcal{F}_1(p'\mathbf n',p\mathbf n)}{\partial p'}\bigg|_{p'=p}\notag \\
&-3 \left(\lambda ^2+1\right) V V ' p  \frac{\partial \mathcal{F}_1(p'\mathbf n',p\mathbf n)}{\partial p'}\bigg|_{p'=p}\notag 
\\
&
+\frac{3}{4} \left[\lambda ^2 \left(V^2-3v^2\right)-2 \lambda  \left(V ^2-v^2\right)+V ^2-v^2\right] \mathcal{F}_1(p\mathbf n',p\mathbf n)
\notag \\
&
+\frac{3}{2} \left(5 \lambda ^2-4 \lambda +2\right) V '{}^2 \mathcal{F}_1(p\mathbf n',p\mathbf n)+3 (\lambda -2) \lambda  V  V ' \mathcal{F}_1(p\mathbf n',p\mathbf n)\bigg],
\\
(n_{\rm e}\sigma_{\rm T}a)^{-1}\mathcal C_{\epsilon^3,0}[f]
=&\int \frac{d\mathbf n'}{4\pi}\frac{T_{\rm e}}{m_{\rm e}}  \Bigg[\frac{3}{4} \left(-11 \lambda ^3+13 \lambda ^2-11 \lambda +9\right)  V ' p^2\frac{\partial^2 \mathcal{F}_1(p'\mathbf n',p\mathbf n)}{\partial p'{}^2}\bigg|_{p'=p}\notag \\
&+\frac{3}{2} \left(2 \lambda ^3-\lambda ^2+2 \lambda -3\right) V  p^2 \frac{\partial^2 \mathcal{F}_1(p'\mathbf n',p\mathbf n)}{\partial p'{}^2}\bigg|_{p'=p}
\notag \\
&+\frac{3}{4} (\lambda -1) \left(\lambda ^2+1\right)  \left(V -V '\right) p^3\frac{\partial^3 \mathcal{F}_1(p'\mathbf n',p\mathbf n)}{\partial p'{}^3}\bigg|_{p'=p}
\notag \\
&
-\frac{3}{8} \left(40 \lambda ^3-47 \lambda ^2+56 \lambda -31\right)  V 'p \frac{\partial \mathcal{F}_1(p'\mathbf n',p\mathbf n)}{\partial p'}\bigg|_{p'=p}\notag \\
&-\frac{3}{8} \left(16 \lambda ^3-41 \lambda ^2+7\right) V  p \frac{\partial \mathcal{F}_1(p'\mathbf n',p\mathbf n)}{\partial p'}\bigg|_{p'=p}\notag\\
&+\frac{3}{4} \left(20 \lambda ^3-46 \lambda ^2+5 \lambda +3\right) V ' \mathcal{F}_1(p\mathbf n',p\mathbf n)\notag \\
&+\frac{3}{8} \left(16 \lambda ^3-41 \lambda ^2+34 \lambda +9\right) V  \mathcal{F}_1(p\mathbf n',p\mathbf n)\Bigg]\notag \\
&\int \frac{d\mathbf n'}{4\pi} \Bigg[
\frac{1}{8} \left(-\lambda ^2-1\right) p^3 \left(V -V '\right) V '{}^2 \frac{\partial^3 \mathcal{F}_1(p'\mathbf n',p\mathbf n)}{\partial p'{}^3}\bigg|_{p'=p}\notag \\
&
+\frac{1}{4} \left(\lambda ^2+1\right) V  p^3 \left(V -V '\right) V ' \frac{\partial^3 \mathcal{F}_1(p'\mathbf n',p\mathbf n)}{\partial p'{}^3}\bigg|_{p'=p}\notag \\
&+\frac{1}{8} \left(-\lambda ^2-1\right) V ^2 p^3 \left(V -V '\right) \frac{\partial^3 \mathcal{F}_1(p'\mathbf n',p\mathbf n)}{\partial p'{}^3}\bigg|_{p'=p}\notag \\
&+\frac{3}{8} \left(\lambda ^2-2 \lambda -1\right) V ^3 p^2 \frac{\partial^2 \mathcal{F}_1(p'\mathbf n',p\mathbf n)}{\partial p'{}^2}\bigg|_{p'=p}\notag \\
&
+\frac{3}{4} \left(3 \lambda ^2-\lambda +2\right) p^2 V '{}^3 \frac{\partial^2 \mathcal{F}_1(p'\mathbf n',p\mathbf n)}{\partial p'{}^2}\bigg|_{p'=p}\notag \\
&
+\frac{3}{8} \left(-11 \lambda ^2+2 \lambda -9\right) V  p^2 V '{}^2 \frac{\partial^2 \mathcal{F}_1(p'\mathbf n',p\mathbf n)}{\partial p'{}^2}\bigg|_{p'=p}\notag\\
&
+\frac{3}{4} \left(2 \lambda ^2+\lambda +3\right) V ^2 p^2 V ' \frac{\partial^2 \mathcal{F}_1(p'\mathbf n',p\mathbf n)}{\partial p'{}^2}\bigg|_{p'=p}\notag \\
&
-\frac{3}{8} V  p \left(\lambda ^2 \left(2 V ^2-7v^{2}\right)-4 \lambda  \left(V ^2-v^{2}\right)+2 V ^2-3v^{2}\right) \frac{\partial \mathcal{F}_1(p'\mathbf n',p\mathbf n)}{\partial p'}\bigg|_{p'=p}\notag \\
&
+\frac{3}{4} \left(15 \lambda ^2-10 \lambda +7\right) p V '{}^3 \frac{\partial \mathcal{F}_1(p'\mathbf n',p\mathbf n)}{\partial p'}\bigg|_{p'=p}-\frac{3}{2} \left(5 \lambda ^2+4\right) V  p V '{}^2 \frac{\partial \mathcal{F}_1(p'\mathbf n',p\mathbf n)}{\partial p'}\bigg|_{p'=p}\notag \\
&
-\frac{3}{8} p \left(\lambda ^2 \left(8 V ^2+7v^{2}\right)-4 \lambda  \left(4 V ^2+v^{2}\right)-4 V ^2+3v^{2}\right) V ' \frac{\partial \mathcal{F}_1(p'\mathbf n',p\mathbf n)}{\partial p'}\bigg|_{p'=p}\notag \\
&
+\frac{3}{8} V  \left(\lambda ^2 \left(2 V ^2-7v^{2}\right)+\lambda  \left(14v^{2}-4 V ^2\right)+2 V ^2-v^{2}\right) \mathcal{F}_1(p\mathbf n',p\mathbf n)\notag \\
&
+3 \left(5 \lambda ^2-5 \lambda +2\right) V '{}^3 \mathcal{F}_1(p\mathbf n',p\mathbf n)+\frac{3}{2} \left(5 \lambda ^2-10 \lambda +2\right) V  V '{}^2 \mathcal{F}_1(p\mathbf n',p\mathbf n)\notag \\
&
+\frac{3}{4} \left(2 \lambda ^2 \left(2 V ^2-7v^{2}\right)+\lambda  \left(13v^{2}-8 V ^2\right)+4 V ^2-5v^{2}\right) V ' \mathcal{F}_1(p\mathbf n',p\mathbf n)
\Bigg].
\end{align}

\subsection{Perturbative expansions}

Next, we integrate the above terms with respect to $\mathbf n'$ and take the harmonic or Legendre coefficients.
(\ref{appen_legen_gattai}) is useful for converting $(\mathbf n\cdot \mathbf n')$ into combinations of $\mathbf n'$ and $\mathbf n$~\footnote{In the previous versions of this paper, we wrongly treated the angular dependence of the perturbations.
The expansion in terms of Legendre polynomial is not possible for the products of perturbations since cosines of Fourier momenta in convolutions are not defined uniquely, and hence we need to expand them by spherical harmonics.}. In this section, we ignore the vector and tensor sector for simplicity and rewrite the physical momenta by using the comoving one.
Here we consider the following perturbative expansion:
\begin{align}
f=\sum_{n=0}^3f^{(n)},~V=\sum_{n=1}^3V^{(n)}\label{expand:pert}
\end{align}
and replace the bulk velocity as $\mathbf v\to \mathbf v^{(1)}+\mathbf v^{(2)}+\mathbf v^{(3)}$.\\

\textit{Zeroth order in the cosmological perturbations---.}
We immediately find that zeroth order collision terms without energy transfer is vanishing:
\begin{align}
(n_{\rm e}\sigma_{\rm T}a)^{-1}\mathcal C^{(0)}_{\rm T}[f]&=0,\label{col:zero}
\end{align}
On the other hand, the Kompaneets terms with $T_{\rm e}/m_{\rm e}$ and $p/m_{\rm e}$ can be written as follows.
\begin{align}
&(n_{\rm e}\sigma_{\rm T}a)^{-1}\mathcal C^{(0)}_{\rm K}[f]=\notag \\
&\frac{ T_{\rm e}}{m_{\rm e}}\left( p^2\frac{\partial^2 f^{(0)}}{\partial p^2}+4 p\frac{\partial f^{(0)}}{\partial p}\right)
+\frac{p(1+z)}{m_{\rm e}}\left(2 f^{(0)}p\frac{\partial f^{(0)}}{\partial p}+p\frac{\partial f^{(0)}}{\partial p}+4f^{(0)}{}^2+4f^{(0)}\right)\label{CK:0}.
\end{align}

\textit{First order in the cosmological perturbations---.}
First order collision term for the Thomson scattering is obtained as
\begin{align}
(n_{\rm e}\sigma_{\rm T}a)^{-1}\mathcal C^{(1)}_{\rm T}[f]&=f^{(1)}_0-f^{(1)}(\mathbf p)-\frac{1}{2}f^{(1)}_2P_2 -  p\frac{\partial f^{(0)}}{\partial p}V^{(1)}.\label{col:first}
\end{align}

We saw that the leading order Kompaneets term is homogeneous, and it is modulated due to the electron density perturbation.
Therefore, introducing the electron density perturbation as
\begin{align}
n_{\rm e}\to n_{\rm e}(1+\delta^{(1)}_{\rm e}),
\end{align}
we have 
\begin{align}
\delta^{(1)}_{\rm e}\mathcal C^{(0)}_{\rm K}[f].
\end{align}

In addition to this effect, the next-to-leading order Kompaneets terms are given as~\cite{Chluba:2012gq}

\begin{align}
&(n_{\rm e}\sigma_{\rm T}a)^{-1}\mathcal C^{(1)}_{\rm K}[f]\notag \\
&=\frac{p(1+z)}{m_{\rm e}} \left[2p \frac{\partial f^{(0)}}{\partial p} f^{(1)}(\mathbf p)+4 f^{(0)} f^{(1)}(\mathbf p)+2 f^{(1)}(\mathbf p)+4 f^{(0)} f^{(1)}_0  +p \frac{\partial f^{(1)}_0}{\partial p}+2 f^{(1)}_0+2 f^{(0)} p\frac{\partial f^{(1)}_0}{\partial p}\right.
\notag\\
&\left.
+(\mathbf{\hat v}\cdot \mathbf n)\left(\frac{24}{5} i f^{(0)} f^{(1)}_1+\frac{12}{5} i f^{(1)}_1+\frac{12i}{5} f^{(0)} p\frac{\partial f^{(1)}_1}{\partial p}+\frac{6 i}{5} p \frac{\partial f^{(1)}_1}{\partial p}\right)\right.\notag \\
&
\left.
+(\mathbf v\cdot \mathbf n) \left(-8 f^{(0)}{}^2-8 f^{(0)}-\frac{7}{5} p^2\frac{\partial^2 f^{(0)}}{\partial p^2}-\frac{14}{5}  f^{(0)} p^2\frac{\partial^2 f^{(0)}}{\partial p^2}-\frac{31}{5} p \frac{\partial f^{(0)}}{\partial p}-\frac{62}{5}  f^{(0)}p \frac{\partial f^{(0)}}{\partial p}\right)\right.\notag \\
&\left.
+P_2 \left(-2 f^{(0)} f^{(1)}_2-f^{(1)}_2- f^{(0)} p\frac{\partial f^{(1)}_2}{\partial p}-\frac{1}{2}p\frac{\partial f^{(1)}_2}{\partial p}\right)\right.\notag \\
&\left.  
+P_3 \left(-\frac{3i}{5}  f^{(1)}_3-\frac{6i}{5} f^{(0)} f^{(1)}_3-\frac{3 i}{5} f^{(0)} p \frac{\partial f^{(1)}_3}{\partial p}-\frac{3 i}{10}  p\frac{\partial f^{(1)}_3}{\partial p}\right)\right]
\notag \\
&
+\frac{T_{\rm e}}{m_{\rm e}} \left[(\mathbf v\cdot \mathbf n) \left(-\frac{7}{5} p^3 \frac{\partial^3 f^{(0)}}{\partial p^3}-\frac{47}{5} p^2 \frac{\partial^2 f^{(0)}}{\partial p^2}-\frac{15}{2} p \frac{\partial f^{(0)}}{\partial p}\right)+\right.\notag \\
&\left.
(\mathbf{\hat v}\cdot \mathbf n)  \left(\frac{6i}{5}  p^2 \frac{\partial^2 f^{(1)}_1}{\partial p^2}+\frac{24i}{5}  p \frac{\partial f^{(1)}_1}{\partial p}+\frac{6i}{5}  f^{(1)}_1\right)+P_2 \left(-\frac{1}{2} p^2 \frac{\partial^2 f^{(1)}_2}{\partial p^2}-2 p \frac{\partial f^{(1)}_2}{\partial p}+3 f^{(1)}_2\right)\right.\notag \\
&\left.
+P_3 \left(-\frac{3i}{10} p^2 \frac{\partial^2 f^{(1)}_3}{\partial p^2}-\frac{6i}{5}  p \frac{\partial f^{(1)}_3}{\partial p}+\frac{6i}{5}  f^{(1)}_3\right)+p^2 \frac{\partial^2 f^{(1)}_0}{\partial p^2}+4 p \frac{\partial f^{(1)}_0}{\partial p}\right]\label{CK:1}.
\end{align}\\

\textit{Second order in the cosmological perturbations---.}
Let us move on to the second order in the cosmological perturbations.
First, the linear collision process above can be modulated because of the fluctuation of electron density. 
This is known as Vishniac effect:
\begin{align}
\delta^{(1)}_{\rm e}\mathcal C^{(1)}_{\rm T}[f].\label{vishniac}
\end{align}
One also finds the following contribution at the next-to-leading order of the energy transfer:
\begin{align}
\delta^{(1)}_{\rm e}\mathcal C^{(1)}_{\rm K}[f],~\delta^{(2)}_{\rm e}\mathcal C^{(0)}_{\rm K}[f].
\end{align}
Another second order contribution is written as follows:

\begin{align}
&(n_{\rm e}\sigma_{\rm T}a)^{-1}\mathcal C^{(2)}_{\rm T}[f]\notag\\
=&f^{(2)}_{0}-f^{(2)}(\mathbf p)+\frac{1}{10}\sum_{m=-2}^{2}Y_{2m}(\mathbf n)f^{(2)}_{2m} \\
&+\frac{1}{10}\sum^{2}_{m=-2}Y_{2m}(\mathbf n) \left(p\frac{\partial}{\partial p} [V^{(1)}f^{(1)}]_{2m} +4 [V^{(1)}f^{(1)}]_{2m}-  p\frac{\partial}{\partial p}  V^{(1)} f^{(1)}_{2m}+  V^{(1)} f^{(1)}_{2m}\right.
\notag \\
&\left.+
\frac{1}{2}[V^{(1)}{}^2]_{2m} p^2  \frac{\partial^2}{\partial p^2}f^{(0)}+5[V^{(1)}{}^2]_{2m} p  \frac{\partial}{\partial p}f^{(0)}
\right)\notag \\
&+\frac{1}{2}\sum^{1}_{m=-1}Y_{1m}(\mathbf n) \left(- V^{(1)} f^{(1)}_{1m}- [V^{(1)}f^{(1)}]_{1m}+V^{(1)}_{1m}f^{(1)}- [V^{(1)}{}^2]_{1m}  p  \frac{\partial}{\partial p}f^{(0)}\right)\notag \\
&+p\frac{\partial}{\partial p}  [V^{(1)}f^{(1)}]_{0}+\frac{5}{2} [V^{(1)}f^{(1)}]_{0}
-p\frac{\partial}{\partial p}  V^{(1)}f^{(1)}_{0}
-p\frac{\partial}{\partial p}  V^{(2)}f^{(0)}
-\frac{1}{2} V^{(1)} f^{(1)}_{0}+\frac{1}{2}V^{(1)}f^{(1)}\notag \\
&+\frac{1}{2}  V^{(1)}{}^2 p^2\frac{\partial^2}{\partial p^2}f^{(0)}+\frac{1}{2} [V^{(1)}{}^2]_{0} p^2  \frac{\partial^2}{\partial p^2}f^{(0)}
+\frac{1}{2} V^{(1)}{}^2 p\frac{\partial}{\partial p}f^{(0)}+\frac{7}{2} [V^{(1)}{}^2]_{0} p  \frac{\partial}{\partial p}f^{(0)}.
\label{trans_sec_col}
\end{align}
We do not expand the products further for notational simplicity, but they can be written by the Gaunt functions.
For example, we can write $[Vf]_{2m}$ as 
\begin{align}
[Vf]_{2m}=v(-1)^{m}\frac{4\pi}{3}\sum_{M'=-1}^{1}\sum_{L=0}^{\infty}\sum_{M=-L}^{L}f_{LM}Y^{*}_{1M'}(\hat{\mathbf v})\mathcal G^{21L}_{-mM'M},
\end{align}
where the Gaunt integral is introduced as
\begin{align}
\mathcal G^{l_{1}l_{2}l_{3}}_{m_{1}m_{2}m_{3}}\equiv \int d\mathbf nY_{l_{1}m_{1}}(\mathbf n)Y_{l_{2}m_{2}}(\mathbf n)Y_{l_{3}m_{3}}(\mathbf n).
\end{align}

\textit{Cubic order in the cosmological perturbations---.}
The electron density modulation also arises at cubic order.
Introducing the cubic order electron density perturbation $\delta^{(3)}_{\rm e}$, we find
\begin{align}
\delta^{(2)}_{\rm e}\left(\mathcal C^{(1)}_{\rm T}+\mathcal C^{(1)}_{\rm K}\right),~\delta^{(1)}_{\rm e}\mathcal C^{(2)}_{\rm T},~\delta^{(3)}_{\rm e}\mathcal C^{(0)}_{\rm K}.
\end{align}
The remaining cubic order Thomson terms are obtained as follows:

\begin{align}
&(n_{\rm e}\sigma_{\rm T}a)^{-1}\mathcal C^{(3)}_{\rm T}[f]\notag \\
=&f_{0}-f(\mathbf p)+\frac{1}{10}\sum_{m=-2}^{2}Y_{2m}(\mathbf n)f_{2m} 
\notag\\
&+\frac{1}{5}\sum^{2}_{m=-2}Y_{2m}(\mathbf n) \left(\frac{1}{2} p\frac{\partial }{\partial p} [Vf]_{2m}
+2 [Vf]_{2m}
-\frac{1}{2} p \frac{\partial }{\partial p}Vf_{2m}+\frac{1}{2} V f_{2m}\right)\notag \\
&+\frac{1}{3}\sum^{1}_{m=-1}Y_{1m}(\mathbf n) \left(-\frac{3}{2} V f_{1m}-\frac{3}{2} [Vf]_{1m}+\frac{3}{2} V_{1m}f\right)\notag \\
&+p \frac{\partial }{\partial p}[Vf]_{0} 
+\frac{5}{2} [Vf]_{0}-V p\frac{\partial f_{0}}{\partial p}-\frac{1}{2} V \left[f_{0}-f\right]
\notag \\
&+\frac{1}{5}\sum^{2}_{m=-2}Y_{2m}(\mathbf n)
 \left(
 \frac{1}{4} p^2 V^2 \frac{\partial^2}{\partial p^2}f_{2m}
 +\frac{1}{4} p^2\frac{\partial^2}{\partial p^2}[V^2 f]_{2m}
 -\frac{1}{2} p^2\frac{\partial^2}{\partial p^2} V [V f]_{2m}-\frac{3}{2} v^2 f_{2m}\right.\notag \\
 &\left.-\frac{1}{2} p\frac{\partial}{\partial p} V^2 f_{2m}+\frac{1}{2} V^2 f_{2m}+\frac{5}{2} p\frac{\partial}{\partial p} [V^2 f]_{2m}-2 p\frac{\partial}{\partial p} V [V f]_{2m}+5 [V^2 f]_{2m}-5 [V^2]_{2m} f
 +2 V [Vf]_{2m}\right)\notag \\
&+\frac{1}{3}\sum^{1}_{m=-1}Y_{1m}(\mathbf n) \left(\frac{3}{2} v^2 f_{1m}+\frac{3}{2}V^2 p  \frac{\partial}{\partial p}f_{1m}-\frac{3}{2} V^2 f_{1m}-\frac{3}{2} p  \frac{\partial}{\partial p}[V^2 f]_{1m}\right.\notag \\
&\left.-6 [V^2f]_{1m}+6 [V^2]_{1m}f-6 V [V f]_{1m}+6 V V_{1m}f\right)\notag \\
&-\frac{3}{2} v^2 \left[f_{0}-f\right]+\frac{1}{2}  V^2 p\frac{\partial}{\partial p}f_{0}+V^2 \left[f_{0}-f\right]+\frac{7}{2} p\frac{\partial}{\partial p}[V^2f]_{0}
\notag \\
&-4  V  p\frac{\partial}{\partial p}[Vf]_{0}+\frac{11}{2} [V^2 f]_{0}-\frac{11}{2} [V^2]_{0}f+V [V f]_{0}\notag \\
&+\frac{1}{2}  V^2 p^2\frac{\partial^2}{\partial p^2}f_{0}+\frac{1}{2} p^2\frac{\partial^2}{\partial p^2}[ V^2 f]_{0}-p^2\frac{\partial^2}{\partial p^2} V [V f]_{0}
\notag \\
&
-\frac{1}{6} V^3  p^3\frac{\partial^3}{\partial p^3}f^{(0)}
+\frac{1}{6} [V^{3}]_{0} p^3\frac{\partial^3}{\partial p^3}f^{(0)} 
-\frac{1}{2} V \left[V^{2}\right]_{0} p^{3}\frac{\partial^3}{\partial p^3}f^{(0)}
\notag \\
&-\frac{1}{4} V^3 p^{2}\frac{\partial^2}{\partial p^2}f^{(0)}
+\frac{9}{4} [V^3]_{0} p^{2}\frac{\partial^2}{\partial p^2}f^{(0)}
-\frac{19}{4} V [V^2]_{0} p^{2}\frac{\partial^2}{\partial p^2}f^{(0)}
\notag \\
&-V^3 p\frac{\partial}{\partial p}f^{(0)}
+9 [V^3]_{0} p\frac{\partial}{\partial p}f^{(0)}-\frac{17}{2} V [V^2]_{0} p \frac{\partial}{\partial p}f^{(0)} +2 v^2 V p\frac{\partial}{\partial p}f^{(0)}\notag \\
&+\frac{1}{3}\sum^{1}_{m=-1}Y_{1m}(\mathbf n) \left(
+6  [V]_{1m} p\frac{\partial}{\partial p}f^{(0)} V^2+\frac{3}{4}[V]_{1m} p^2  \frac{\partial^2}{\partial p^2}f^{(0)} V^2
+\frac{3}{4}  [V^2]_{1m} p^{2}\frac{\partial^2}{\partial p^2}f^{(0)} V\right.\notag \\
&\left.
-\frac{15}{2}  [V^3]_{1m} p\frac{\partial}{\partial p}f^{(0)}+\frac{3}{2}  v^2 [V]_{1m} p\frac{\partial}{\partial p}f^{(0)}-\frac{3}{4}  [V^3]_{1m} p^{2}\frac{\partial^2}{\partial p^2}f^{(0)}\right)\notag \\
&+\frac{1}{5}\sum_{m=-2}^{2}Y_{2m}(\mathbf n) \left(
\frac{1}{12} [V^3]_{2m} p^{3} \frac{\partial^3}{\partial p^3}f^{(0)}
-\frac{1}{4} V [V^2]_{2m} p^{3}\frac{\partial^3}{\partial p^3}f^{(0)}
+\frac{3}{2} [V^3]_{2m} p^{2}\frac{\partial^2}{\partial p^2}f^{(0)}\right.\notag \\
&\left.-\frac{11}{4} V [V^2]_{2m}p^{2} \frac{\partial^2}{\partial p^2}f^{(0)}
+\frac{15}{2} [V^3]_{2m} p \frac{\partial}{\partial p}f^{(0)}
-5 V [V^2]_{2m} p \frac{\partial}{\partial p}f^{(0)}
\right),\label{col:cubic_thomson}
\end{align}
where we did not use Eq.~(\ref{expand:pert}) to simplify the expression.
This expression is complicated, but its frequency dependence is classified into three types as we discuss in the following sections.

\section{Solutions to the Boltzmann equation}\label{sec:3}

The cosmic microwave background radiation is almost an isotropic and ideal blackbody at high precision but slightly deviates from the perfect one.
In this section, we construct such a deviation as a solution of the second order Boltzmann equation.

\subsection{Momentum functions}\label{U:e}

We start with introducing the following functional basis to expand the distribution function:

\begin{align}
\mathcal G(p)&=-p\frac{\partial f^{(0)}(p)}{\partial p},\label{def:G}\\
\mathcal Y(p)&=\frac{1}{p^2}\frac{\partial }{\partial p}p^4\frac{\partial f^{(0)}(p)}{\partial p}, \label{def:Y}\\
\mathcal M(p)&=T_0\frac{\partial f^{(0)}(p)}{\partial p},\label{def:M}
\end{align}
where we have defined
\begin{align}
f^{(0)}(p)=\frac{1}{e^{\frac{p}{T_0}}-1}.
\end{align}
The definite integrals of these functions are
\begin{align}
\int^\infty_0 \frac{dx}{2\pi^2}x^n f^{(0)}&=\mathcal I_n,\label{B:2}\\
\int^\infty_0 \frac{dx}{2\pi^2}x^n \mathcal G&=(n+1)\mathcal I_n,
\\
\int^\infty_0 \frac{dx}{2\pi^2}x^n \mathcal Y&=(n+1)(n-2)\mathcal I_n,\\
\int^\infty_0 \frac{dx}{2\pi^2}x^n \mathcal M&=-n\mathcal I_{n-1},
\end{align}
where $x=p/T_0$ and $\{I_n\}_{n=1}^3=\{1/12,\zeta(3)/\pi^2,\pi^2/30\}$.
The following relations are also useful:

\begin{align}
p^2\frac{\partial^2 f^{(0)}(p)}{\partial p^2}&=\mathcal Y+4\mathcal G,\label{2.9}\\
p\frac{\partial}{\partial p}p\frac{\partial f^{(0)}(p)}{\partial p}&=\mathcal Y+3\mathcal G\label{pdelp}\\
-p\frac{\partial f^{(1)}(p)}{\partial p}&=\Theta^{(1)}(3\mathcal G+\mathcal Y).\label{3.4}
\end{align}
More generally,
\begin{align}
p^n\frac{\partial }{\partial p}p^m\frac{\partial}{\partial p}p^lf^{(0)}=p^{n+m+l-2}\left[l(l+m-1)f^{(0)}+\mathcal Y+(4-m-2l)\mathcal G\right].
\end{align}
These $\mathcal G$, $\mathcal Y$ and $\mathcal M$ are ``linearly independent''.
Let us consider a linear combination of these functions, which is equal to zero:
\begin{align}
a \mathcal G+ b\mathcal Y+ c \mathcal M=0. 
\end{align}
We then integrate the both side with respect to momentum $p$ and obtain 
\begin{align}
a(n+1)\mathcal I_n+b(n+1)(n-2)\mathcal I_n-cn\mathcal I_{n-1}=0.\label{proof:linear}
\end{align}
The solution to (\ref{proof:linear}) can be found to be trivial so that the $\mathcal G$, $\mathcal Y$ and $\mathcal M$ are linearly independent.

\subsection{First order}

The cosmic photon fluid perturbs along the primordial fluctuations; however, it is considered to be in thermal equilibrium at each point.
Based on the assumption, we usually write the ansatz for the first order Boltzmann equation as a Planck distribution function with a spacetime dependent temperature.
Let us expand the distribution function up to the linear order in terms of the temperature perturbations.
Using (\ref{def:G}), we can write the linear term as
\begin{align}
f^{(1)}=\Theta^{(1)} \mathcal G,\label{linear_ansatz}
\end{align}
where we have defined $\Theta^{(1)}$ as the first order temperature fluctuations normalized by the fiducial temperature.
One then finds that both sides of the Boltzmann equation are proportional to $\mathcal G$.
This implies that the equation for $\Theta^{(1)}$ coincides with those of the energy density perturbations and the number density perturbations, which are integrated with respect to the momentum.
At this stage, we can justify the first assumption that the system is blackbody locally.

\subsection{Second order in the Thomson limit}

On the other hand, it has been already known that the above discussions are not applicable at second order~\cite{Dodelson:1993xz}.
In other words, the second~(or higher) order temperature perturbations are momentum dependent in general, and the fluid is no more a local blackbody.
This is apparent if we integrate the Boltzmann equation with $p^n$  and find different equations for the number density and the energy density at second order.
We have three strategies to solve this Boltzmann equation.
One is to integrate the momentum as we did right now and to focus on the energy density or the number density.
It can be a significant simplification, at the same time, it masks much information in the non-trivial momentum dependence of the distribution function.
The second way is to consider the full momentum dependent temperature.
This can be perfect, but it is far more complicated.
We then propose to take into account the momentum dependence partly by the form of the spectral distortions.
In other words, we replace the infinite number of degrees of freedom coming from the continuous frequency with the infinite number of the parameters describing the spectral deformations.
In fact, we usually apply this kind of approach to reduce the number of equations of a set of partial differential equations.
For example, we use the Boltzmann hierarchy equations instead of the equations with angular parameters.
In this case, the crucial contributions are related to the lower multipoles, and the infinite number of equations fall into a few equations.
We have already made this approach even for our problem at linear order.
We can say that the first order spectral distortion is written as momentum independent temperature perturbations in the form of (\ref{linear_ansatz}), that is, an infinite number of d.o.f of continuous momentum is reduced to a single local parameter at first order.
Our next step is going to second or higher order.\\

Let us write an arbitrary distribution function in the form of the Planck distribution function with momentum dependent temperature perturbation $\widetilde \Theta=\widetilde \Theta(\mathbf x, p\mathbf n, \eta)$:

\begin{align}
f(\mathbf x, p\mathbf n, \eta)&=\frac{1}{e^{\frac{p}{T_0}e^{-\widetilde\Theta}}-1},
\end{align}
where we define $T_0$ as the temperature of a time independent comoving blackbody~\footnote{$T_0$ is not the average temperature which varies due to the acoustic reheating or other heating processes at second order.}, 
$\mathbf x$ is the comoving spacial coordinate, and $\eta$ is the conformal time hereafter.
Then let us expand this function around $\widetilde \Theta=0$
\begin{align}
\frac{1}{e^{\frac{p}{T_0}e^{-\widetilde\Theta}}-1}=\sum^{\infty}_{n=0}\frac{\widetilde \Theta^n}{n!}\left(-p\frac{\partial }{\partial p}\right)^n f^{(0)}(p).\label{planck_dist_exp}
\end{align}
In our convention, the zeroth order distribution function is time independent, and it simplifies the later calculations.
Here we should notice that the $n$ is not the order of the perturbations since the temperature perturbations have the following form:

\begin{align}
\widetilde \Theta=\sum_{m} \widetilde \Theta^{(m)},
\end{align}
where $m$ is the order of the perturbations.
We have already known that $\widetilde \Theta^{(1)}=\Theta^{(1)}$ is momentum independent; however, 
$n>1$ is momentum dependent  in principle.
In our definition, the higher order temperature perturbations have none zero homogeneous component since we fix the fiducial temperature $T_0$ as mentioned above.
At second order, we find that the momentum dependence is separated as
\begin{align}
\widetilde \Theta^{(2)}(p)=\Theta^{(2)}+y\frac{\mathcal Y(p)}{\mathcal G(p)},
\end{align}
where $\Theta^{(2)}$ is the momentum independent part and $\mathcal Y$ is defined in (\ref{def:Y}).
The perturbative expansions are obtained as follows:
\begin{align}
f^{(1)}=&\Theta^{(1)} \mathcal G(p),\label{def:f1}\\
f^{(2)}=&\left[\Theta^{(2)}+\frac32\Theta^2\right] \mathcal G(p) + \left[y+\frac12\Theta^2\right]\mathcal Y(p)\label{def:f2},
\end{align}
where we have used (\ref{def:G}) and (\ref{def:Y}).

\subsection{Boltzmann equations}

Thanks to the expression (\ref{def:f2}), it simplifies calculations to write the second order collision terms as

\begin{align}
\mathcal C^{(2)}_{\rm T}[f]+\delta^{(1)}_{\rm e}C^{(1)}_{\rm T}[f]=\mathcal A^{(2)}\mathcal G+\mathcal B^{(2)}\mathcal Y,\label{decomp:2nd}
\end{align}
where we have used (\ref{vishniac}) and (\ref{trans_sec_col}) with (\ref{def:f2}).
Thus, frequency dependence at second order falls into $\mathcal G$ and $\mathcal Y$.

We will solve the Boltzmann equation perturbatively; however, note that we avoid writing the second order metric perturbations explicitly in the following discussions since they do not appear in the final expressions for the spectral distortions.

So far we have discussed the r.h.s. of the Boltzmann equation.
Next, let us see the Liouville term on the left.
Differentiating the distribution function with respect to the conformal time, (\ref{def:f2}) yields
\begin{align}
f'=&f^{(0)'}+(\Theta'+3\Theta\Theta')\mathcal G+
\left(\Theta+\frac32\Theta^2\right)\mathcal G'\notag \\
&+(y'+\Theta\Theta')\mathcal Y+\left(y+\frac12\Theta^2\right)\mathcal Y',\label{fprime}
\end{align}
where $'\equiv d/d\eta$.
Since $f^{(0)}$ depends only on comoving momentum $p$, the derivative of the zeroth order part becomes
\begin{align}
f^{(0)'}=- (\ln p)'\mathcal G.\label{f0prime}
\end{align}
Note that $- (\ln p)'$ does not have the zeroth order part but have both the first and the second order terms which describe the gravitational redshift since $p$ is the comoving momentum.
On the other hand, using (\ref{pdelp}) we can write the time derivative of $\mathcal G$ by
\begin{align}
\mathcal G'=-(\ln p)'(3\mathcal G+\mathcal Y),\label{Gprime}
\end{align}
and $\mathcal Y'$ can be neglected since this is the first order quantity which is multiplied by the second order perturbations in the equation.
Combining (\ref{fprime}), (\ref{f0prime}) and (\ref{Gprime}) up to second order, we obtain
\begin{align}
f'=(1+3\Theta)[\Theta'-(\ln p)']\mathcal G+\left( y'+
\Theta[\Theta'-(\ln p)']\right)\mathcal Y.\label{fprimeGY}
\end{align}
%


Next, let us write down equations order by order.
The first order Boltzmann equation is easily obtained as 

\begin{align}
\Theta^{(1)'}-(\ln p)'^{(1)}&= \mathcal A^{(1)}.\label{bol:1st}
\end{align}
If we expand $(\ln p)'^{(1)}$ with respect to the metric perturbations, we obtain first order Boltzmann equation of the temperature perturbations with metric perturbations.
On the other hand, collecting second order terms, the second order equation can be written as 

\begin{align}
\left[\Theta^{(2)'}-(\ln p)'^{(2)}+3\Theta\mathcal A^{(1)}\right]\mathcal G+
\left[y'+\Theta\mathcal A^{(1)}\right]\mathcal Y
&= \mathcal A^{(2)}\mathcal G +\mathcal B^{(2)}\mathcal Y,\label{bol:2nd}
\end{align}
where we have used (\ref{bol:1st}) for substituting $\mathcal A^{(1)}$ into the expression.
$\mathcal A^{(2)}$ does not have any monopole terms since the Compton scattering does not change the photon number; however, the temperature is raised by acoustic reheating, that is, the spacial configuration is changed due to $3\Theta \mathcal A^{(1)}$ on the left.
Integrals of the Boltzmann equation with respect to $p^n$ should always be consistent even if they do not have any physical implications since we should respect the equation at the distribution function level.
Therefore, each coefficient for $\mathcal G$ and $\mathcal Y$ should be equal so that we obtain the Boltzmann equations for $\Theta^{(2)}$ and $y$   independently.
One then immediately finds
\begin{align}
\left[\Theta^{(2)'}-(\ln p)'^{(2)}+3\Theta\mathcal A^{(1)}\right]&=\mathcal A^{(2)},\label{sec:eq}\\
y'+\Theta\mathcal A^{(1)}=\mathcal B^{(2)}.\label{y:eq}
\end{align}

In contrast to (\ref{bol:1st}), there are source terms in the l.h.s. of (\ref{sec:eq}), and this implies that the small-scale perturbation generates large-scale temperature perturbations at second order, whose homogenous component is recently pointed out in~\cite{Jeong:2014gna,Nakama:2014vla}.
It is crucial that $\mathcal B^{(2)}$ not have any pure second-order terms except $y$.
If one does not introduce the $y$ to the distribution function at the beginning, (\ref{bol:2nd}) is not satisfied since both $\Theta\mathcal A^{(1)}$ and $\mathcal B^{(2)}$ are already determined at first order and do not coincide in principle.
This also implies that the $y$ is determined by products of the linear perturbations automatically.
Let us expand (\ref{y:eq}) by substituting the following form:

\begin{align}
\mathcal A^{(1)}&=n_{\rm e}\sigma_{\rm T} a\left(\Theta_0-\Theta +V+\frac{1}{10}\sum_{m=-2}^{2}Y_{2m}(\mathbf n)\Theta_{2m}\right),\\
\mathcal B^{(2)}&=n_{\rm e}\sigma_{\rm T} a\bigg[y_{0} -y
+\frac{1}{10}\sum_{m=-2}^{2}Y_{2m}(\mathbf n)y_{2m}\notag \\
&
+\frac{1}{2}[\Theta^{2}]_{0} -\frac12\Theta^{2}
+\frac{1}{20}\sum_{m=-2}^{2}Y_{2m}(\mathbf n)[\Theta^{2}]_{2m}
+V\Theta_{0}-[V\Theta]_{0}+\frac{1}{2}V^2 
+\frac{1}{2}[V^2]_{0}\notag \\
&+\frac{1}{10}\sum_{m=-2}^{2}Y_{2m}(\mathbf n)\left[V\Theta_{2m} -[V\Theta]_{2m}+\frac12[V^2]_{2m}\right] \bigg ].
\end{align}

Then we find
\begin{align}
\frac{\partial y}{\partial\eta}+\mathbf n\cdot \nabla y&=n_{\rm e}\sigma_{\rm T} a \left(\Theta -\Theta_0-V-\frac{1}{10}\sum_{m=-2}^{2}Y_{2m}(\mathbf n)\Theta_{2m}\right)\Theta\notag \\
& +n_{\rm e}\sigma_{\rm T} a\bigg[y_{0} -y
+\frac{1}{10}\sum_{m=-2}^{2}Y_{2m}(\mathbf n)y_{2m}\notag \\
&
+\frac{1}{2}[\Theta^{2}]_{0} -\frac12\Theta^{2}
+\frac{1}{20}\sum_{m=-2}^{2}Y_{2m}(\mathbf n)[\Theta^{2}]_{2m}
+V\Theta_{0}-[V\Theta]_{0}+\frac{1}{2}V^2 
+\frac{1}{2}[V^2]_{0}\notag \\
&+\frac{1}{10}\sum_{m=-2}^{2}Y_{2m}(\mathbf n)\left[V\Theta_{2m} -[V\Theta]_{2m}+\frac12[V^2]_{2m}\right] \bigg ].\label{y:evol:eq}
\end{align}
The equations for the spectral distortions do not include the other pure second order quantities such as the temperature and the metric perturbations.
Therefore, we do not have to integrate the full second-order Boltzmann equation as long as working only on the spectral distortions. 
In Fourier space, the convolutions include the curvature perturbations implicitly as discussed in appendix \ref{app_conv}.
Therefore, the integration with respect to the Fourier momentum is non-trivial in general.

\section{Inhomogeneous $y$ distortion}\label{sec:4}

We have introduced $y$ distortion as momentum dependent part of the second order temperature perturbations.
Its momentum dependence is the same form with the usual Compton $y$ parameter given as
\begin{align}
y_{\rm C}= \int  \frac{T_{\rm e}-T_{\gamma}}{m_{\rm e}}n_{\rm e}\sigma_{\rm T}ad\eta,\label{def:yc}
\end{align}
where $T_{\gamma}\equiv T_0(1+z)$.
We should note that our $y$ is a free parameter determined by the second order Boltzmann equation and has nothing to do with the inhomogeneity of $T_{\rm e}$, $T_{\gamma}$ and $n_{\rm e}$ in the integrand in (\ref{def:yc}).
To be more specific, our $y$ arises from the expansion associated with the Thomson collision terms but $y_{\rm C}$ comes from the Kompaneets terms, that is, their momentum dependences coincide accidentally. 
Below we summarize the evolution equation for our $y$ and confirm availability of the previous estimations.

\subsection{Generation of the monopole spectral distortion}

For the long wavelength limit, we obtain the following equation for the monopole component of the $y$ distortion:

\begin{align}
( n_{\rm e}\sigma_{\rm T} a)^{-1}\dot y_{0}\approx & [(\Theta-V)^{2}]_{0}-\Theta_{0}^{2}-\frac{1}{4\pi\cdot 10}\sum_{m=-2}^{2}\Theta_{2m}\Theta_{2,-m}(-1)^{m},
\end{align}
where $\cdot \equiv\partial/\partial\eta$.
The $l>0$ linear perturbations are not significant before the horizon entry. 
Therefore, using (\ref{conv_result1}), each convolution should be well approximated as
\begin{align}
(XY)_{\bf k}\sim \int \frac{dq}{q}X_{q} Y_{q}\mathcal P_{\mathcal R}(q)\mathcal R^{(2)}_{\mathbf k}.\label{convo:sub}
\end{align}
(\ref{convo:sub}) implies that the source terms induce $k$ independent transfer functions for the $y$ distortion on large scales.
Ignoring the gradient terms, the transfer function for the monopole $y$ distortion follows
\begin{align}
\dot y_0 \sim& n_{\rm e}\sigma_{\rm T} a  \int \frac{dq}{q}\mathcal P_{\mathcal R}(q) \left(\frac{1}{3}(iv-3\Theta_{1})^{2}+\frac{9}{2} \Theta_2^2+\cdots\right ),\label{S_0}
\end{align}
where dots imply the higher order multipole components.
The first term in r.h.s. of (\ref{S_0}) is heat conduction from baryons, and the second one is shear viscosity.
For the higher order multipoles, we have $- n_{\rm e}\sigma_{\rm T} ay_l$, which suppresses growing of $y_l$ by isotropization due to the Thomson scattering.
The main part of $y_0(\eta,k)$ is entirely the same form with the homogeneous component previously evaluated, for example, in~\cite{Chluba:2012gq}.
This is reasonable since the monopole and homogeneous part are not distinguishable before horizon entry.

\subsection{Integral solutions and Gauge dependence}

Let us demonstrate a line-of-sight integral method for the $y$ distortion.
Near the last scattering surface, the source can be negligible and the equations in Fourier space can be written as
\begin{align}
\dot y+ik\lambda y&=-n_{\rm e}\sigma_{\rm T} a\left(y-y_{0}+\frac12P_2(\lambda)y_2\right).\label{y:equation:fourier}
\end{align}
We do this calculation in a completely parallel way with that for the temperature perturbations.
That is, the line-of-sight integral solution for the $y$ distortion is given by
\begin{align}
y(k,\lambda,\eta_0)=&\int^{\eta_0}_{\eta_f}d\eta \mathcal S_y(k,\eta)e^{-ik(\eta_0-\eta)\lambda}\\
\mathcal S_y(k,\eta)=&g\left(y_0+\frac{y_2}{4}+\frac{3\ddot y_2}{4k^2}\right)+\dot g\frac{3\dot y_2}{2k^2}+\ddot g\frac{3y_2}{4k^2},
\end{align}
where the visibility function is introduced by $g=-\dot \tau e^{-\tau}$ with $\dot \tau$ being $-n_{\rm e}\sigma_{\rm T}a$.
The terms related to $y_2$ are new corrections.
The harmonic coefficient is also immediately obtained as

\begin{align}
a_{y,lm}&=4\pi(-i)^l\int\frac{d^3k}{(2\pi)^3}Y^*_{lm}(\hat k)\int^{\eta_0}_0 d\eta \mathcal S_y(k,\eta)j_l[k(\eta_0-\eta)].\label{yualm}
\end{align}

We can see that no metric perturbations and no second order temperature perturbations are included.
Therefore, $y$ has no redshift and no cross-correlation with the ISW lensing. 
This helps us to consider the $\mu\mu$ and $yy$ auto and $\mu y$ cross-correlations since we do not have to consider the curve of sight~\cite{Saito:2014bxa}.\\

At the end of this section, we comment on the gauge dependence of $y$.
The gauge transformation laws for $v$ and $-3i\Theta_1$ are the same as given in~\cite{Ma:1995ey}, and higher order multipoles are gauge invariant quantities; therefore, gauge invariance of $y$ is manifest~\cite{Chluba:2012gq}.
There is no metric perturbation in (\ref{y:equation:fourier}) so that $y$ distortion evolves gauge independently after its generation.

\section{$\mu$ distortion}\label{sec:5}

\subsection{Definition}

We have shown that the $y$ is necessary for a set of equations to be consistent at second order; however, we have not commented on the chemical potential type distortion called $\mu$ distortion.
During $5\times 10^4<z<2\times 10^6$, the $y$ distortions are converted to the $\mu$ distortion, and the system is considered to be in kinetic equilibrium.
This was investigated numerically in the previous studies~\cite{Hu:1994bz,Chluba:2012gq}.
Let us try to include the $\mu$ as well in our formulation.

One strategy to include the chemical potential may be writing a second order ansatz in the following form: 
\begin{align}
\tilde \Theta^{(2)}(p)=\Theta^{(2)}+y\frac{\mathcal Y(p)}{\mathcal G(p)}+\mu\frac{\mathcal M(p)}{\mathcal G(p)}+\cdots,\label{ans:muiri}
\end{align}
where $\mathcal M$ is defined in (\ref{def:M}).
Let us substitute (\ref{ans:muiri}) into (\ref{trans_sec_col}) and (\ref{fprimeGY}).
We then obtain additional terms proportional to $\mathcal M$.
Reading off each coefficient, the evolution equation for the $\mu$ distortion is given as follows:
\begin{align}
\dot \mu +ik\lambda \mu&=(n_{\rm e}\sigma_{\rm T}  a)\left[-\mu+\mu_{0}+\frac{1}{10}\sum_{m=-2}^2\mu_{2m}Y_{2m}\right].\label{mu_eq}
\end{align}
It is not surprising that the conversion of $y$ to $\mu$ is not seen even at second order since we start with a momentum independent $\mu$ parameter, and we did not take into account the momentum transfer.
(\ref{mu_eq}) just tells us that the momentum independent chemical potential evolves independently from the $y$ distortion and the second order temperature perturbations once it is given at initial time.
This implies that $\mu$ generation from $y$ should be treated in the full~(or higher order) Boltzmann equations with momentum dependent chemical potential.
The other steps for the $\mu$ are completely parallel with those for the $y$, and the harmonic coefficient is the same form with (\ref{yualm}), that is, we have

\begin{align}
a_{\mu,lm}=&4\pi(-i)^l\int\frac{d^3k}{(2\pi)^3}Y^*_{lm}(\hat k)\int^{\eta_0}_0 d\eta \mathcal S_\mu(k,\eta)j_l[k(\eta_0-\eta)],\label{mualm}\\
\mathcal S_\mu(k,\eta)=&g\left(\mu_0+\frac{\mu_2}{4}+\frac{3\ddot \mu_2}{4k^2}\right)+\dot g\frac{3\dot \mu_2}{2k^2}+\ddot g\frac{3\mu_2}{4k^2},
\end{align}
and the initial value of the $\mu$ distortion should be introduced by hands in this context.

\subsection{Instantaneous $\mu$ formation}

The full numerical analysis for $\mu$ generation is complicated.
Here we notice that the chemical potential is a thermodynamic quantity and that we do not have to care about the details of the process when considering the thermalization timescale is rapid enough.
In this section, we repeat a traditional explanation for the $\mu$ formation with a single comment.
Let us consider that the initial state is given by the solution of the Thomson limit second order Boltzmann equation, that is, the second order number and the energy density are calculated as
\begin{align}
N^{(2)}_y&=0,\\
I^{(2)}_y&=4y\mathcal I_3,
\end{align}
where numerical factors $\mathcal I_{n}$ are defined in (\ref{B:2}).
Assuming that the thermalization timescale is rapid enough compared to the typical timescale of the cosmic expansion, these quantities should have the following forms at the next moment:
\begin{align}
N^{(2)}_{\rm BE}&=3\mathcal I_2\Theta^{(2)}_{\rm BE}-2\mu \mathcal I_1\\
I^{(2)}_{\rm BE}&=4\mathcal I_3\Theta^{(2)}_{\rm BE}-3\mu \mathcal I_2
\end{align}
where subscript ``BE'' implies that they are the parameters associated with a Bose distribution function. 
Then we can impose the number and the energy conservation laws:
\begin{align}
N^{(2)}_y&=N^{(2)}_{\rm BE}\label{N_equate}\\
I^{(2)}_y&=I^{(2)}_{\rm BE}\label{I_equate}
\end{align}
so that we obtain
\begin{align}
\mu=\left(\frac{2\mathcal I_1}{3\mathcal I_2}-\frac{3\mathcal I_2}{4\mathcal I_3}\right)^{-1}y.\label{e:mu}
\end{align}
The numerical constant is calculated as $\mu=1.40066\times 4y$, and the well known relation is derived.
Now we have a comment on this matter.
(\ref{N_equate}) and (\ref{I_equate}) should not be established at a distribution function level, that is, this approach never explains the continuous evolution of the $\mu$.
This is because the momentum integrals with $p^n$ should always be consistent if we start with the Boltzmann equation.
Suppose that, for instance, we use the Boltzmann equations which are integrated with $p^4$ and $p^5$, we find the other numerical factor in (\ref{e:mu}).
Therefore, there exist time discontinuities in both sides of the equalities in (\ref{N_equate}) and (\ref{I_equate}).
Using (\ref{S_0}) and (\ref{e:mu}) we approximately obtain the following form of $\mu$ distortion from the scalar perturbations:

\begin{align}
\mu_0(\eta_f,k)\sim \int^{\eta_f}_{\eta_i}  n_{\rm e}\sigma_{\rm T} a d\eta\int \frac{dq}{q}\mathcal P_{\mathcal R}(q) \left(\frac{1}{3}(iv-3\Theta_{1})^{2}+\frac{9}{2} \Theta_2^2+\cdots\right ).\label{mu0:calc:def}
\end{align}

\subsection{Monopole formation}

We have mentioned that the $\mu$ formation is non-trivial and our prescription is not applicable.
Here let us revisit the monopole terms in (\ref{CK:1}).
These terms coincide with those in (\ref{CK:0}).
This implies that the previous numerical simulation based on (\ref{CK:0}) is also applicable to the inhomogeneous case as long as we ignore the higher order multipoles.
As discussed in the previous section, multipoles of the spectral distortions are vanishing, and only the long wavelength modes of the monopole component are dominant.
In this sense, we expect that the generation of the inhomogeneous distortions should be explained in the same manner as the previous numerical calculations for the homogeneous $\mu$ distortions~\cite{Hu:1994bz,Chluba:2012gq}.

\subsection{Suppression from the Double Compton scattering}

So far we have introduced the initial redshift for the $\mu$ by referring to the previous numerical works~\cite{Hu:1994bz,Chluba:2012gq}.
It is determined by the double Compton effect, which is cubic order QED interaction.
The process is crucial for the spectral distortions since it changes the number of photons and erases the distortions.
The derivation of the double Compton scattering collision term is complicated so that we avoid writing the term explicitly here.
Instead, roughly we estimate the timescales of these interactions.
Let $\Gamma_{\rm K}$ and $\Gamma_{\rm DC}$ be the energy transfer ratio due to the Compton scattering and double Compton scattering, respectively. 
$\Gamma_{\rm K}$ should be proportional to the scattering event ratio $n_{\rm e}\sigma_{\rm T}$.
Note that there is no energy transfer only if $n_{\rm e}\sigma_{\rm T}$ is large. 
There exist significant energy transfers if electrons are relativistic enough to transfer the photon energy.
This should be characterized by $T_{\rm e}/m_{\rm e}$.
Then, we can estimate the energy transfer ratio due to the Compton scattering as follows:
\begin{align}
\Gamma_{\rm K}\sim n_\text{e}\sigma_{\rm T}\frac{T_\gamma}{m_{\rm e}}.
\end{align}
Actually, we can reproduce this relation by using (\ref{CK:0}).
In analogy with the above discussion, we can roughly estimate $\Gamma_{\rm DC}$ as well.
First, the double Compton scattering interaction ratio should be proportional to $n_{\rm e}\sigma_{\rm T}\alpha \epsilon^2$ with $\alpha$ being the fine structure constant.
This is because the process is a cubic order QED interaction, and $\epsilon \to0$ limit electron does not emit the second photon in terms of energy conservation law~\footnote{
Linear terms in $\epsilon$ do not exist since the scattering cross section should be a Lorentz scalar.}.
Then, the lowest order term can be estimated as
\begin{align}
\Gamma_{\rm DC}\sim n_\text{e}\sigma_T\alpha\left(\frac{T_{\gamma}}{m_\text{e}}\right)^2.\label{timsecale_dc}
\end{align}
Using $\Gamma_{\rm K}$ and $\Gamma_{\rm DC}$, we can guess the suppression time scale of the $\mu$ distortion.
The $\mu$ distortion varies as a result of both the double Compton scattering and the Compton scattering.
Therefore, the suppression time scale can be given as the inverse of $\Gamma_\mu\sim \sqrt{\Gamma_{\rm DC}\Gamma_{\rm K}}$.
Employing these facts, we roughly obtain $\Gamma_{\mu}\sim 10^{-35}\times(1+z)^{\frac92} {\rm s}^{-1}$ and $\Gamma_{\rm K}\sim 10^{-29}\times(1+z)^{4} {\rm s}^{-1}$.
Comparing these with $H\sim 10^{-20}\times(1+z)^{2}{\rm s}^{-1}$, one finds that the window of $\mu$ era is opened during $\mathcal O(10^5)<z<\mathcal O(10^6)$.

\section{Higher order spectral distortions}\label{sec:6}

A product of distribution functions in (\ref{dist:expand}) can be expanded as
\begin{align}
&g(\mathbf{\tilde q}')f(\mathbf {\tilde p}')[1+f(\mathbf {\tilde p})]-g(\mathbf{\tilde q})f(\mathbf{\tilde  p})[1+f(\mathbf{\tilde p}')]\notag \\
&=g(\mathbf{\tilde q})\left(f(\mathbf {\tilde p}')-f(\mathbf {\tilde p})+\left(-\frac{(\mathbf {\tilde p}-\mathbf {\tilde p}')^2}{2m_{\rm e}T_{\rm e}}-\frac{(\mathbf{\tilde q}-m_{\rm e} \mathbf v)\cdot (\mathbf {\tilde p}-\mathbf {\tilde p}')}{m_{\rm e}T_{\rm e}}\right)f(\mathbf {\tilde p}')\left[1+f(\mathbf {\tilde p})\right]+\cdots \right).\notag\\
&=g(\mathbf{\tilde q})\left[f(\mathbf {\tilde p}')-f(\mathbf {\tilde p})+\mathcal O\left(\frac{\eta}{\epsilon}\right)\right].
\end{align}
This implies that the collision terms are linear in $f$ if we ignore the momentum transfer corrections coming from $\tilde p/m_{\rm e}$ and $T_{\rm e}/m_{\rm e}$.
We start this section with the above Thomson scattering limit.

\subsection{Cubic order ansatz at Thomson limit}

\begin{figure}
\includegraphics[width=15cm]{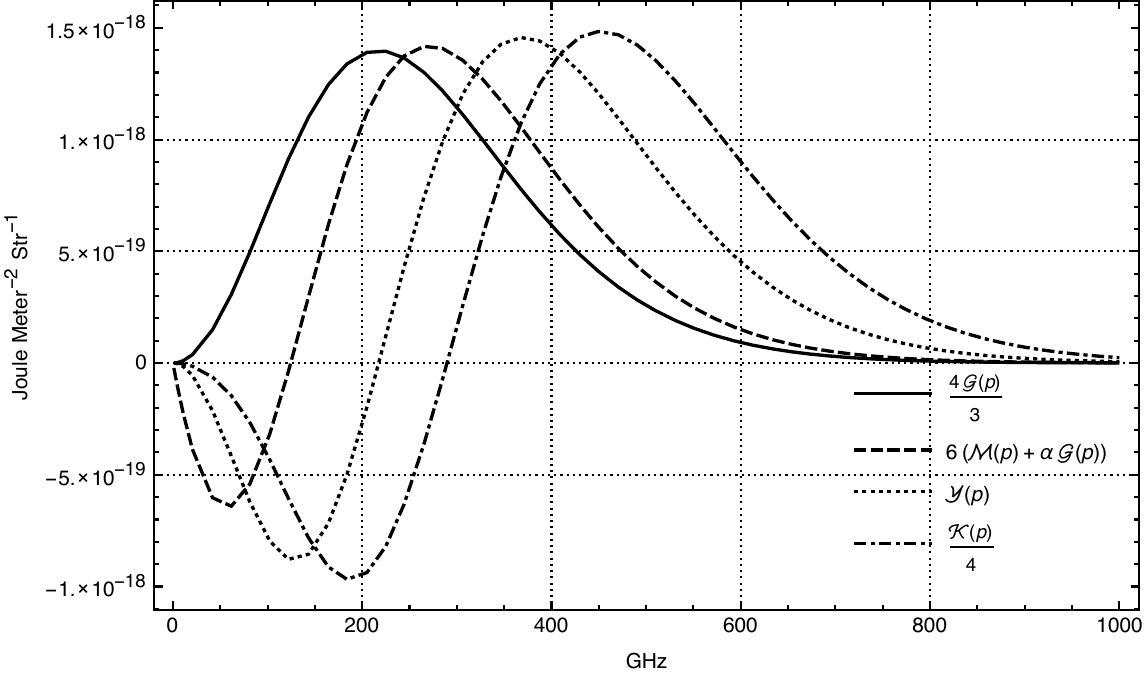}
\caption{Spectral shapes of the photon number shift, $\mu$ distortion, $y$ distortion and higher order $y$ distortion are drawn.
They are rescaled for comparing the shapes and the peaks.
The multiples are shown in the legend in the figure.}
\label{fig_1}
\end{figure}

The dimensional quantity is only the photon momentum in the collision terms.
Therefore, the derivative operators always appear in the form of $p\partial/\partial p$.
The cubic order Thomson term $(n_{\rm e}\sigma_{\rm T}a)^{-1}\mathcal C^{(3)}_{\rm T}[f]$ should be written as a linear combination of
\begin{align}
f,\quad  p\frac{\partial f}{\partial p},\quad  \left(p\frac{\partial }{\partial p}\right)^2f ,\quad \left(p\frac{\partial }{\partial p}\right)^3 f,\label{dim:an}
\end{align}
and their Legendre coefficients with the baryon bulk velocity as explicitly shown in (\ref{col:cubic_thomson}).
Let us introduce a following momentum function: 
\begin{align}
\mathcal K(p)=\left(-p\frac{\partial }{\partial p}\right)\mathcal Y(p),
\end{align}
where the momentum integral of $\mathcal K$ with $p^2$ is 0, which inspires us to define a higher order $y$ distortion. 
Using this function, the third order derivative of the Planck distribution is given as  
\begin{align}
\left(-p\frac{\partial }{\partial p}\right)^3 f^{(0)}(p)&=\mathcal K +3\mathcal Y+9\mathcal G.
\end{align}
Combining the above with (\ref{planck_dist_exp}), one finds the following cubic order terms:

\begin{align}
f^{(3)}=\widetilde \Theta^{(3)}\mathcal G+ \widetilde \Theta^{(1)}\widetilde \Theta^{(2)}\left(3\mathcal G + \mathcal Y\right)
+\frac{\widetilde\Theta^{(1)3}}{3!}\left(9\mathcal G + 3 \mathcal Y+ \mathcal K\right).\label{cubic_expand}
\end{align}
Then we separate the momentum dependence of the temperature perturbations as
\begin{align}
\widetilde \Theta^{(1)}(p)&=\Theta^{(1)}\\
\widetilde \Theta^{(2)}(p)&=\Theta^{(2)}+\frac{\mathcal Y}{\mathcal G}y^{(2)}\\
\widetilde \Theta^{(3)}(p)&=\Theta^{(3)}-\frac{\mathcal Y^2}{\mathcal G^2}\Theta^{(1)} y^{(2)}+\frac{\mathcal Y}{\mathcal G}y^{(3)}+\frac{\mathcal K}{\mathcal G}\kappa^{(3)},
\end{align}
and we can write the cubic order terms as
\begin{align}
f^{(3)}&=\left[\Theta^{(3)}+3\Theta^{(1)}\Theta^{(2)}+\frac32\Theta^{(1)3}\right]\mathcal G\notag\\
&+\left[\Theta^{(1)}\Theta^{(2)}+\frac12\Theta^{(1)3}+3\Theta^{(1)}y^{(2)}+y^{(3)}\right]\mathcal Y\notag \\
&+\left[\frac1{3!}\Theta^{(1)3}+\kappa^{(3)} \right]\mathcal K.\label{cubic_ans}
\end{align}
This is our ansatz for the cubic order Thomson limit Boltzmann equation.
The above discussion suggests that closed equations for the higher order spectral distortions such as $\Theta^{(3)}$, $y^{(3)}$ and $\kappa^{(3)}$ are systematically obtained.
We can reconstruct the distribution functions in the form of the sum of local blackbody and spectral distortions.
(\ref{cubic_expand}) and (\ref{cubic_ans}) yield  
\begin{align}
f=\frac{1}{e^{\frac{p}{T_0}e^{-\Theta}}-1}+\left[(1+3\Theta^{(1)})y^{(2)}+y^{(3)}\right]\mathcal Y+\kappa^{(3)}\mathcal K,
\end{align}
where we have defined momentum independent temperature perturbation as $\Theta = \Theta^{(1)}+\Theta^{(2)}+\Theta^{(3)}$.
The spectral shapes of the momentum basis are shown in Fig.\ref{fig_1}.
Defining $\alpha=2\mathcal I_1/(3\mathcal I_2)$, conventionally the $\mu$ distortion is expressed by not $\mathcal M$ but $\mathcal M+ \alpha \mathcal G$, which is the difference between a Bose and Planck distributions whose number densities are the same.
$y^{(3)}$ can be subdominant part of $y^{(2)}$; however, we can distinguish $\kappa^{(3)}$ from these due to the momentum dependence even if its magnitude is smaller than $y$ distortions.

\subsection{General ansatz at Thomson limit}

The same prescription is available for higher orders as long as we assume the linearity of the distribution functions.
Let us introduce $n$-th order momentum function whose integral with $p^2$ is zero:
\begin{align}
\mathcal Y^{(n+1)}(p)=\left(-p\frac{\partial }{\partial p}\right)^n\mathcal Y(p),
\end{align}
where $\mathcal Y^{(1)}=\mathcal Y$ and $\mathcal Y^{(2)}=\mathcal K$.
Then, $l$-th term in (\ref{planck_dist_exp}) is expressed as
\begin{align}
\left(-p\frac{\partial }{\partial p}\right)^l f^{(0)}(p)&=\mathcal Y^{(l-1)}+3\mathcal Y^{(l-2)}+\cdots +3^{l-2}\mathcal Y^{(1)}+3^{l-1}\mathcal G\\
&=3^{l-1}\mathcal G(p)+\sum^{l-1}_{k=1}3^{l-k-1}\mathcal Y^{(k)}(p).
\end{align}
As discussed in (\ref{dim:an}), the momentum dependence is always expressed by linear combination of $\mathcal G$, $\mathcal Y^{(1)},\cdots$ and $\mathcal Y^{(n-1)}$.
Using these functions, the $n$-th order distribution function should be written as
\begin{align}
f^{(n)}=\left[\Theta^{(n)}+\cdots\right]\mathcal G +\cdots +\left[\cdots +y^{(n-1,1)}\right]\mathcal Y^{(n-2)}+\left[\frac{1}{n!}\left(\Theta^{(1)}\right)^n+y^{(n,0)}\right]\mathcal Y^{(n-1)},
\end{align}
where $y^{(2)}=y^{(2,0)}$, $y^{(3)}=y^{(2,1)}$ and $\kappa^{(3)}=y^{(3,0)}$.
A number of the new parameters for the $n$-th order Thomson limit Boltzmann equations can be $n$.
On the other hand, the time derivative of the momentum basis is calculated as
\begin{align}
\mathcal Y^{(n)'}=-(\ln p)'\mathcal Y^{(n+1)}.
\end{align}
Using this with the same manipulation for (\ref{fprimeGY}), acoustic sources for higher order distortions can be written as 
\begin{align}
y^{(n,0)'}=-\frac{1}{(n-1)!}{\Theta^{(1)}}^{n-1}\mathcal A^{(1)}+\cdots.
\end{align}
Therefore, we always have the higher order spectral distortions as results of mode couplings as in the case with the usual $y$ distortion.

\subsection{First order Kompaneets terms}

So far we have discussed the Thomson limit to ignore the nonlinear terms of $f$ for simplicity.
The above prescription itself can be powerful since it is applicable for the same class of collision processes; however, we should take into account not only the inhomogeneity but also the momentum transfer in the practical application to the CMB.
When we discussed the second order theory, the Kompaneets terms are comparable to the Thomson anisotropic parts; however, they are homogenous and do not contribute to the perturbation equation.
The total average part is calculated by combining the result of the Thomson part with the SZ effects.
If we look at the cubic order, the momentum transfer is expected to be written as products of the Compton $y$ parameter and the first order anisotropies.
In this case, the linear Kompaneets terms are non-negligible for the perturbation equations.

We now discuss the momentum transfer coming from $p(1+z)/m_{\rm e}$ and $T_{\rm e}/m_{\rm e}$ at cubic order.
From (\ref{CK:0}) and (\ref{CK:1}), we have 
\begin{align}
(n_{\rm e}\sigma_{\rm T}a)^{-1}\left(\mathcal C^{(0)}_{\rm K,0}[f]+\mathcal C^{(1)}_{\rm K,0}[f]\right)=\frac{1}{m_{\rm e}p^2}\frac{\partial }{\partial p}p^4\left(T_{\rm e}(1+{\Theta_{\rm e0}})\frac{\partial f_0}{\partial p}+f_0\left[1+f_0\right]\right),\label{617}
\end{align}
where $f_0=f^{(0)}+f^{(1)}_0$, and we replace $T_{\rm e}\to T_{\rm e}(1+{\Theta_{\rm e0}})$ to include the electron temperature perturbation.
The differentiated part can be calculated as
\begin{align}
T_{\rm e}(1+{\Theta_{\rm e0}})\frac{\partial f_0}{\partial p}+f_0\left[1+f_0\right]=\left[T_{\rm e}(1+\Theta_{\rm e0})-T_0 e^{\widetilde\Theta_0}\right]\frac{\partial f_0}{\partial p}-\frac{T_0e^{\widetilde\Theta_0}}{1-p\frac{\partial \widetilde\Theta_0}{\partial p}}\frac{\partial \widetilde\Theta_0}{\partial p}p\frac{\partial f_0}{\partial p}.\label{f0:kernel_Komp}
\end{align}
Therefore, (\ref{617}) yeilds
\begin{align}
(n_{\rm e}\sigma_{\rm T}a)^{-1}\mathcal C^{(1)}_{\rm K,0}[f]\simeq\frac{1}{m_{\rm e}p^2}\frac{\partial}{\partial p}p^4\left[T_{\rm e}(1+\Theta^{(1)}_{\rm e0})-T_{\gamma} (1+\Theta^{(1)}_0)\right]\frac{\partial f_0}{\partial p}.\label{f0:kernel_Komp}
\end{align}
The momentum independence of $\Theta^{(1)}$ is important for this expression, and we have nontrivial additional terms at higher order.
Then, the monopole component of the first order Kompaneets equation is obtained as follows~\footnote{The terms proportional to $v$ do not contain the zeroth order distribution functions so that they are the second order Kompaneets terms.}:

\begin{align}
(n_{\rm e}\sigma_{\rm T}a)^{-1}\mathcal C^{(1)}_{\rm K,0}[f]&=\frac{1}{m_{\rm e}p^2}\frac{\partial}{\partial p}p^4\left[(T_{\rm e}\Theta_{\rm e0}-T_{\gamma}\Theta_0)\frac{\partial f^{(0)}}{\partial p}+
(T_{\rm e}-T_{\gamma})\Theta \frac{\partial \mathcal G}{\partial p}\right]\notag \\
&=\frac{T_{\rm e}\Theta_{\rm e0}-T_{\gamma}\Theta_0}{m_{\rm e}}\mathcal Y+\frac{T_{\rm e}-T_{\gamma} }{m_{\rm e}} \Theta_0 \mathcal K,\label{1st:Komp}
\end{align}
where we use
\begin{align}
\frac{1}{p^2}\frac{\partial }{\partial p}p^4\frac{\partial }{\partial p}=-3\left(-p\frac{\partial }{\partial p}\right)+\left(-p\frac{\partial }{\partial p}\right)^2,
\end{align}
and 
\begin{align}
\frac{1}{p^2}\frac{\partial }{\partial p}p^4\frac{\partial }{\partial p}\mathcal G=\mathcal K.
\end{align}
The other cubic order terms should be linear combinations of $\mathcal G$, $\mathcal Y$ and $\mathcal K$ as pointed above.

\subsection{Linear Sunyaev-Zel'dovich effect}

(\ref{1st:Komp}) has terms proportional to $\mathcal Y$ and $\mathcal K$.
The first term implies that there are additional sources for (\ref{y:eq}).
Assuming that $T_{\rm e}=T_{\gamma}$,
\begin{align}
\frac{T_{\rm e}\Theta_{\rm e0}-T_{\gamma}\Theta_0}{m_{\rm e}}=\frac{T_{\gamma}}{3m_{\rm e}}S_{\rm e \gamma},\label{iso_sz}
\end{align}
where we have defined the baryon isocurvature perturbation as
\begin{align}
S_{\rm e \gamma}=3(\Theta_{\rm e0}-\Theta_0)
. 
\end{align}
This implies that the $yT$ cross correlation function does exist even for Gaussian perturbations suppose that there are baryon isocurvature perturbations and that they are cross correlated with the adiabatic ones.
Physical implication of (\ref{iso_sz}) is clear: the fluctuations of relative number density induce additional recoil effects.
These terms may be crucial since $T_{\gamma}/m_{\rm e}=\mathcal O(10^{-9})(1+z)$, which may be comparable to the acoustic source for $z\gtrsim \mathcal O(10^3)$.

On the other hand, for the adiabatic initial condition $\Theta_{\rm e0}=\Theta_0$, one finds
\begin{align}
n_{\rm e}\sigma_{\rm T}a\frac{T_{\rm e}\Theta_{\rm e0}-T_{\gamma}\Theta_0}{m_{\rm e}}=\dot y_{\rm C}\Theta_0.
\end{align}
This should be more important since $y_{\rm C}$ is recently estimated in~\cite{Hill:2015tqa}, and the magnitude is expected to be $10^{-6}$.
Therefore, we roughly expect
\begin{align}
y^{(3)}_0\sim \int d\eta \dot y_{\rm C} \Theta_0\sim y_{\rm C} \Theta_0(z_c), 
\end{align}
where $z_c$ is the redshift when SZ effects occur.
The cross correlation with the temperature can be given as
\begin{align}
C^{y^{(3)} T}_{l}\sim 10^{-6}C_l^{TT}.\label{3yt}
\end{align}
If we compare (\ref{3yt}) to the non-Gaussianity origin $y^{(2)}T$ cross correlation, this corresponds to $f^{\rm loc}_{\rm NL}\sim 10$~\cite{Chluba:2016aln}, that is, we cannot ignore this contribution for the non-Gaussianity observation by using $yT$ cross correlation; however, we also point out that the systematic errors for the PIXIE experiment is $10^3$ times larger than the signal~\cite{Kogut:2011xw} and the problem is not so simple.

The second term in (\ref{1st:Komp}) also implies the other higher order SZ effects.
As in the case with $y^{(3)}$, we estimate the higher order spectral distortion as
\begin{align}
\kappa^{(3)}_0\sim y_{\rm C}\Theta_0(z_c).
\end{align}
This implies that there are two types of linear SZ effects, and we can distinguish this higher order distortion from the former tSZ effect due to the momentum dependence.
We should note that the anisotropies in the distortions are connected with the electron gas configurations, and it may be possible to have a 3D map of the linear perturbations by using the linear SZ effects.

\subsection{Higher order spectral distortions and a residual distortion}

Recently, non-$\mu$ and non-$y$ type spectral distortions called \textit{residual distortion} is proposed for the purpose of classifying the actual observational data of CMB intensity spectrum~\cite{Chluba:2013pya}.
The authors introduced the $n$ dimensional Euclidean space with $n$ being many frequency channels and characterized the distortions by using linearly independent vectors in the space.
The residual distortion is the fourth direction perpendicular to temperature shift, $y$ distortion, and $\mu$ distortion directions.
There are $n-3$ linearly independent directions for residual distortion, and our higher order momentum basis $\mathcal Y^{(n)}$ should be included in them.
The residual distortion mainly well describes thermal history during $\mu$-$y$ transition period, which cannot be treated in our method due to the non-linearity of the distribution functions in Kompaneets terms as seen in~(\ref{f0:kernel_Komp}).
Full parametrization of the residual distortion in a systematic approach should be necessary for the future of observational cosmology.

\section{Summary}

The second order temperature perturbations are momentum dependent in contrast to zeroth and first order.
The momentum is usually integrated to obtain the second order brightness perturbations so that non-trivial configurations in the momentum spectrum has not been analyzed.
In this paper, we explicitly wrote the second order Boltzmann equation for the Planck distribution function with a momentum dependent temperature and showed that such dependence is separated into two ``linearly independent'' functions with corresponding parameters.
One of them is understood as the fluctuations of the local blackbodies. 
Another is the form of well known $y$ distortion which arises from Silk damping.
We derived the exact evolution equation of the distortion and combined it with the homogeneous component coming from the other thermal history.
On the other hand, we also showed that the formation of the spectral $\mu$ distortion is not understood in our framework.
The $\mu$ is a result of the frequent momentum transfer, and the momentum independent ansatz does not work to explain the generation.
In the last section, we also discussed the potential to extend our method to a higher order.
In our cases, the linearity of the distribution functions in the collision terms is crucial.
As an example, we investigated the cubic order Boltzmann equations.
We derived the cubic order Thomson terms and linear Kompaneets terms, and newly define the higher order $y$ distortion to make the equations closed.
We also showed that the mode coupling arises as in the case with $y$ distortions, and found linear SZ effects. 
The above method has a potential for applying to broader classes of non-equilibrium physics or non-linear problems.
The basis functions may have the different forms depending on concrete collision terms; however several classes may be solved as we have shown in this paper.
For example, the Maxwell-Boltzmann distribution functions with several orders of the spectral distortions may be another window for the analysis of large-scale structure, and Boltzmann equations for the massive neutrino might be solved in the same manner.

\acknowledgments

We also would like to thank Enrico Pajer, Jens Chluba and Wayne Hu for a lot of helpful discussions.
We would like to thank Masahide Yamaguchi for a lot of helpful comments.
We would like to thank Jens Chluba, Keisuke Inomata and Taku Haga for careful reading of our manuscript.
The author is supported by a Grant-in-Aid for the Japan Society for the Promotion of Science~(JSPS) Fellows.

\appendix

\section{Multipole and harmonic expansion}
In this paper, the multipole expansion of $X$ is defined as 

\begin{align}
X(\hat{\mathbf v}\cdot \mathbf n)&=\sum_{l} (-i)^l(2l+1)P_l(\hat{\mathbf v}\cdot\mathbf n)X_l \label{Legendre},
\end{align}
where $\hat {\mathbf v}$ is the direction of the baryon bulk velocity.
For the linear perturbation, $\hat{\mathbf v}$ is parallel to Fourier momentum.
On the other hand, the harmonic expansion is introduced as
\begin{align}
X(\mathbf n)&=\sum_{l,m}X_{lm}Y_{lm}(\mathbf n).
\end{align}
The relation between these coefficients for $m=0$ can be expressed as
\begin{align}
X_{l0}=\sqrt{4\pi(2l+1)}(-i)^lX_l.
\end{align}
For example, expanding (\ref{def:f1}) and (\ref{def:f2}) in terms of multipoles, we can write the coefficients as follows:
\begin{align}
f^{(1)}_l&=\Theta^{(1)}_l \mathcal G,\\
f^{(2)}_l&=\left(\Theta^{(2)}_l+\frac32[\Theta^2]_l\right) \mathcal G + \left(y_l+\frac12[\Theta^2]_l\right)\mathcal Y
\end{align}
Products of $(\mathbf n\cdot \mathbf n')$ can be separated by using the following formula:
\begin{align}
P_{l}(\mathbf n\cdot \mathbf n')=
\frac{4\pi}{2l+1}\sum^{l}_{m=-l}Y_{lm}(\mathbf n)Y^{*}_{lm}(\mathbf n').\label{appen_legen_gattai}
\end{align}

\section{A translation of second order collision terms}\label{appendix:second_def}

While we reproduce the second order Boltzmann collision terms which was derived in a previous literature, we comment on the correspondence of the variables with the result in ~\cite{Bartolo:2006cu}.
(4.42) in~\cite{Bartolo:2006cu} is written by 
\begin{align}
\frac12\mathcal C[f](n_e\sigma_T)^{-1}=&\frac12f^{(2)}_{00}-\frac14\sum_{m=-2}^{2}\frac{\sqrt{4\pi}}{5^{3/2}}f^{(2)}_{2m}Y_{2m}-\frac12f^{(2)}(\bm p)\notag \\
&
+\delta_e^{(1)}\left[f^{(1)}_0+\frac12f_2^{(1)}P_2(\mathbf {\hat v}\cdot \mathbf n)-f^{(1)}-p\frac{\partial f^{(0)}}{\partial p}(\mathbf v\cdot \mathbf n)\right]\notag \\
&
-\frac12p\frac{\partial f^{(0)}}{\partial p}({\bf v}^{(2)}\cdot \mathbf n)\notag \\
&
+(\mathbf v\cdot \mathbf n)
\left[f^{(1)}(\bm p)-f^{(1)}_0-p\frac{\partial f^{(1)}_0}{\partial p}-f^{(1)}_2
+\frac12P_2(\mathbf{\hat v}\cdot \mathbf n)\left(f^{(1)}_2-p\frac{\partial f^{(1)}_2}{\partial p}\right)
\right]
\notag \\
&+v\left[2f^{(1)}_1+p\frac{\partial f^{(1)}_1}{\partial p}+\frac{1}{5}P_2(\hat{\mathbf v}\cdot \mathbf n)\left( -f^{(1)}_1(p)+p\frac{\partial f^{(1)}_1}{\partial p}+6f^{(1)}_3+\frac32p\frac{\partial f^{(1)}_3}{\partial p}\right)\right]
\notag \\
&+(\mathbf v\cdot\mathbf n)^2\left[p\frac{\partial f^{(0)}}{\partial p}+\frac{11}{20}p^2\frac{\partial^2 f^{(0)}}{\partial p^2}\right]+v^2\left[p\frac{\partial f^{(0)}}{\partial p}+\frac{3}{20}p^2\frac{\partial^2 f^{(0)}}{\partial p^2}\right]
\notag \\
&+\frac{1}{m_{\rm e}p^2}\frac{\partial}{\partial p}\left[p^4\left(T_{\rm e}\frac{\partial f^{(0)}}{\partial p}+f^{(0)}(1+f^{(0)})\right)\right],
\end{align}
where we found $p^2$ in the denominator of the last line as also pointed out in~\cite{Senatore:2008vi}.
In our notation, the time coordinate is conformal time $\eta$ and we replace the functions as
\begin{align}
\mathbf p&\to \mathbf{\tilde p},\\
\frac12\mathcal C[f]&\to \mathcal C[f],\\
\frac12f^{(2)}&\to f^{(2)},\\
\frac12\mathbf v^{(2)}&\to \mathbf v^{(2)}\\
f_l&\to (-i)^lf_l,\\
f_{lm}&\to (-i)^{-l}\sqrt{\frac{2l+1}{4\pi}}f_{lm}.
\end{align}

\section{A treatment to convolutions}\label{app_conv}

The linear perturbations are merely proportional to the primordial perturbations in Fourier space.
Therefore, the Boltzmann equations are also the equations for the transfer functions simultaneously.
This is not the case at second order since we have convolutions, that is, the Fourier momentum integrals which include the primordial curvature perturbations.
Let us write the curvature perturbations in the convolution explicitly as follows:
\begin{align}
(XY)_{\mathbf k}\equiv \int \frac{d^3 q}{(2\pi)^3}X_{q}Y_{|\mathbf k-\mathbf q|}\mathcal R_{\mathbf q}\mathcal R_{\mathbf k-\mathbf q},
\end{align}
where let $X_{q}$ and $Y_{|\mathbf k-\mathbf q|}$ be the transfer functions for the linear perturbations.
The ensemble average with $\mathcal R_{\mathbf k'}$ is then given as
\begin{align}
\langle (XY)_{\mathbf k} \mathcal R_{\mathbf k'}\rangle =(2\pi)^3\delta^{(3)}(\mathbf k+\mathbf k')\int \frac{d^3 q}{(2\pi)^3}X_{q}Y_{|\mathbf k-\mathbf q|}B_{\mathcal R}(q,|\mathbf k-\mathbf q|,k'),
\end{align}
where $B_{\mathcal R}$ is the shape function of the primordial bispectrum.
Let us consider the case $X$, and $Y$ are significant for $k\ll q$, and let us integrate $q$ in advance.
Then, we can approximate the above equation as
\begin{align}
\langle (XY)_{\mathbf k} \mathcal R_{\mathbf k'}\rangle \simeq(2\pi)^3\delta^{(3)}(\mathbf k+\mathbf k')\int \frac{d^3 q}{(2\pi)^3}X_{q}Y_{q}B_{\mathcal R}(q,q,k').
\end{align}
Suppose that the bispectrum is local-type, one can simplify this further and obtain
\begin{align}
\langle (XY)_{\mathbf k} \mathcal R_{\mathbf k'}\rangle =(2\pi)^3\delta^{(3)}(\mathbf k+\mathbf k')P_{\mathcal R}(k')\left(-\frac{12}{5}f^{\rm loc}_{\rm NL}\right)
\int \frac{dq}{q}\mathcal P_{\mathcal R}(q)X_{q}Y_{q}.
\end{align}
This expression tells us that it is equivalent to replace the convolution as 
\begin{align}
(XY)_{\mathbf k}\sim \int \frac{dq}{q}\mathcal P_{\mathcal R}(q)X_{q}Y_{q}\mathcal R^{(2)}_{\mathbf k},\label{conv_result1}
\end{align}
where we have defined the ``second order curvature perturbation'' to satisfy 
\begin{align}
\langle \mathcal R^{(2)}_{\mathbf k}\mathcal R_{\mathbf k'}\rangle &\sim (2\pi)^3\delta^{(3)}(\mathbf k+\mathbf k')\left(-\frac{12}{5}f^{\rm loc}_{\rm NL}\right)P_{\mathcal R}(k),\\
\langle \mathcal R^{(2)}_{\mathbf k}\mathcal R^{(2)}_{\mathbf k'}\rangle &\sim (2\pi)^3\delta^{(3)}(\mathbf k+\mathbf k')4\tau^{\rm loc}_{\rm NL}P_{\mathcal R}(k).
\end{align}

\bibliography{12bib}{}
\bibliographystyle{unsrt}

\end{document}